\nonstopmode
\documentclass[usenatbib]{mn2e}
\usepackage{graphicx}
\usepackage[mathscr]{euscript}
\usepackage{cuted}
\usepackage{hyperref}
\hypersetup{colorlinks, citecolor = [rgb]{0,0,1} }
\usepackage{amsmath}
\usepackage{soul}

\newcommand{\vect}[1]{\boldsymbol{#1}}
\def\gsim{\mathrel{\raise.3ex\hbox{$>$\kern-.75em\lower1ex\hbox{$\sim$}}}}
\def\lsim{\mathrel{\raise.3ex\hbox{$<$\kern-.75em\lower1ex\hbox{$\sim$}}}}
\def\leq{\mathrel{\raise.3ex\hbox{$<$\kern-.75em\lower1ex\hbox{$-$}}}}

\newcommand{\ie}{\textit{i}.\textit{e}.}
\newcommand{\eg}{\textit{e}.\textit{g}.}

\makeatletter
\newcommand\footnoteref[1]{\protected@xdef\@thefnmark{\ref{#1}}\@footnotemark}
\makeatother

\begin{document}

\title{The Peculiar Velocity Correlation Function}
\author[Wang, Rooney, Feldman \& Watkins]{Yuyu Wang$^{1,\star}$, Christopher Rooney$^{2}$, Hume A.Feldman$^{1}$ \& Richard Watkins$^{3}$\\
$^1$Department of Physics \& Astronomy, University of Kansas, Lawrence, KS 66045, USA.\\
$^2$Department of Astronomy, Cornell University, Ithaca, NY 14850, USA.\\
$^3$Department of Physics, Willamette University, Salem, OR 97301, USA.\\
E-mails: $^\star$yuyuwang@ku.edu}
\maketitle

\begin{abstract}
We present an analysis of the two-point peculiar velocity correlation function using data from the CosmicFlows catalogues.  The Millennium and MultiDark Planck 2 N-body simulations are used to estimate cosmic variance and uncertainties due to measurement errors.  We compare the velocity correlation function to expectations from linear theory to constrain  cosmological parameters. Using the maximum likelihood method, we find values of $\Omega_m= 0.315^{+0.205}_{-0.135}$ and $\sigma_8=0.92^{+0.440}_{-0.295}$, consistent with the Planck and Wilkinson Microwave Anisotropy Probe CMB derived estimates. However, we find that the cosmic variance of the correlation function is large and non-Gaussian distributed, making the peculiar velocity correlation function less than ideal as a probe of large-scale structure.  
\end{abstract}


\section{INTRODUCTION}
\label{sec:introduction}

The peculiar velocity field is a sensitive probe of mass fluctuations on large scales and a powerful tool for constraining cosmological parameters. However, the precision measurement of the velocity field is limited by the error in the measurement of radial distance. Many methods have been introduced to measure the distance with the smallest possible error, such as Tully-Fisher (TF) \citep{TullyFisher1977}, Faber-Jackson \citep{FaberJackson1976}, and the Fundamental Plane (FP) \citep{DjoDav1987,DreLynBurDav1987}. 

These methods do not directly measure radial distance. Rather they estimate the distance modulus, which is proportional to the logarithm of the distance. While the errors in the distance modulus are Gaussian, the distances themselves have a non-Gaussian error distribution which may bias the results.  To address this issue, \citet{WatFel2015a} introduced a new unbiased estimator of the peculiar velocity that gives Gaussian distributed errors. In this paper we will use this unbiased estimator together with the measured redshift \citep[see also][]{DavScr2014,TulCouSor2016} to derive the velocity correlation function.

The fractional observational errors of the radial distances are typically of the order of $\approx 20$\% \citep[e.g.][]{MasSprHay2006, SprMasHay2007, TulCouDol2013}, and peculiar velocities tend to have errors proportional to the distances, which may be large. Because of this large error, a single peculiar velocity measurement is not a good approximation of the velocity of a galaxy. However, statistical ensembles, especially the low-order moment statistics, may be a good estimator of the cosmic velocity field and thus a good tracer of the underlying mass distribution in the Universe \citep[e.g.][]{FelWat2008,WatFelHud2009,FelWatHud2010, DavNusMas2011,NusBraDav2011,MacFelFer2011,Turnbull2012,MacFelFer2012,Nusser2014,SprMagColMou2014,JohBlaKod2014,ScrDavBlaSta2015}.

Many recent studies have focused on the bulk flow, which is the lowest order statistic of the velocity field and is generally thought of as the average of peculiar velocities in a volume \citep[e.g.][]{AbaFel2012,Nusser2014,KumWanFelWat2015,SeiPar2016,ScrDavBla2015,Nusser2016}.  Bulk flows are typically calculated using one of two popular methods.

The first is the maximum likelihood estimate (MLE) method \citep[e.g.][]{Kaiser1988,WatFel2007}. The MLE formalism estimates the bulk flow as a weighted average of the sample velocities, with the weights calculated to minimize its overall uncertainty given the positions, velocities and errors distributions in the catalogue.
The formalism reduces the entire data set to three numbers, namely the components of the bulk flow vector. Since the particular data and error distribution in the surveys analysed are unique to each catalogue, it is difficult to compare the bulk flow calculated using this method between independent surveys. 

The other popular formulation is the minimum variance (MV) method \citep{WatFelHud2009, FelWatHud2010, ScrDavBlaSta2015}.
The MV formalism minimizes the differences between the actual observational data and an 'ideal' survey that may be designed to probe a volume in a particular way. It can be used with Gaussian-weighted \citep{AgaFelWat2012} or tophat-weighted ideal survey distributions \citep{DavNusMas2011,HofNusCorTul2016}.  Because it uses a standard ideal survey bulk flow as a reference, it easily lends itself to direct comparisons between independent surveys. 

Another approach to studying the large-scale velocity field is the pairwise velocity statistic ($v_{12}$) \citep{FerJusFel1999, JusFerFelJaf2000, FelJusFer2003, Hellwing2014, HelBarFre2014}, which is the mean value of the peculiar velocity difference of a galaxy pair at separation $\vect{r}$. Recent studies show that it can also be used to detect the kinetic Sunyaev-Zeldovich effect \citep{ZhaFelJus2008, HanAddAub2012, PlanckXXXVII2015}. 

In this paper we will use a different approach to probe the cosmic velocity field, namely the peculiar velocity correlation function. It was first introduced by \citet{Gorski1988} and further elucidated in \citet{GorDavStr1989}. 
In subsequent studies, the velocity correlation function has shown potential for providing interesting constraints on cosmological parameters 
\citep[e.g.][]{JafKai1995,ZarZehDekHof1997,JusFerFelJaf2000,BorCosZeh2000, AbaErd2009, NusDav2011, OkuSelVla2014, HowStaBla2017}.  At the time that the original studies of velocity correlation were done, the small sizes of peculiar velocity catalogues limited its usefulness.   The recent availability of large, calibrated catalogues of peculiar velocities and large-scale cosmological simulations suggests that it is  worthwhile to revisit the velocity correlation function as a cosmological probe.   

Here, we will present a feasibility study of this statistic for the study of the large-scale-structure given state-of-the-art peculiar velocity catalogues CosmicFlow-2 (CF2) \citep{TulCouDol2013} and CosmicFlows-3 (CF3) \citep{TulCouSor2016}.   In particular, we assess the magnitude of the cosmic variance expected in the correlation function using mock catalogues extracted from the Millennium Simulation\footnote{\label{note1}http://gavo.mpa-garching.mpg.de/Millennium/}.   Using this result, we show that the velocity correlation functions calculated from CF2 and CF3 are consistent with the standard cosmological model. However,  our results suggest that the particular cosmological volume we live in is on the higher end of the cosmic variance, suggesting that the $\sim$150$h^{-1}$Mpc radius region around us has greater large-scale motions than one would expect on average.   

The organization of this paper is as follows: in section~\ref{sec:catalogue}, we detail our use of N-body simulations to generate mock surveys. In section~\ref{sec:correlation_function}, we introduce the velocity correlation statistic and the methods used to calculate it. In Section~\ref{sec:error}, we discuss the use of the Monte Carlo method for error analysis. In section~\ref{sec:CR}, we show the results for the velocity correlation function using the CosmicFlows catalogues. In Section~\ref{sec:linear}, we explore using the correlation function to constrain cosmological models. Section~\ref{sec:conclusion} concludes this paper.

\section{Mock Catalogues}
\label{sec:catalogue}

To study the properties of the velocity correlation function, we use mock catalogues generated from the Virgo - Millennium Database\footnoteref{note1} of the Millennium Simulation \citep{Millennium1}, which contains the result of the L-Galaxies run used in \citet{DeLBla2007}. The Millennium Simulation is a dark matter only simulation using the GADGET-2 simulation code \citep{Millennium1}. Table~\ref{T_Mill} \citep{GuoWhiAng2013} shows the cosmological parameters of the simulation we used.
We generate 100 mock catalogues for each real survey centred at random locations in the box. Each mock catalogue is designed to have the same radial selection function as the real CF2/CF3 surveys we use here.

\begin{table}
\caption{The cosmological parameters of the Millennium Simulation}
\centering
{
\begin{tabular}{lc}
\hline
Matter density, $\Omega_m$ & 0.272\\
Cosmological constant density, $\Omega_\Lambda$ & 0.728\\
Baryon density, $\Omega_b$ & 0.045\\
Hubble parameter, $h$ ($100 km s^{-1} Mpc^{-1}$) & 0.704\\
Amplitude of matter density fluctuations, $\sigma_8$ & 0.807\\
Primordial scalar spectral index, $n_s$ & 0.967\\
Box size ($h^{-1}$Mpc) & 500\\
Number of particles & $2160^3$\\
Particle mass, $m_p$ ($10^{8} h^{-1} M_{\odot}$) & 8.61\\
Softening, $f_c$ ($h^{-1}kpc$) & 5\\
\hline
\end{tabular}%
}
\label{T_Mill}
\end{table}

Each of the CosmicFlows catalogues comes in two versions, one where galaxies are given individually, which we will call the galaxy compilation, and a group catalog where galaxies in known groups have had their distance moduli and redshifts averaged, resulting in a single velocity and position for the group as a whole .   The CF2 galaxy catalogue (CF2-galaxy)  \citep{TulCouDol2013} contains 8135 galaxies, whereas the CF2 group catalogue (CF2-group) contains 4842 galaxies and groups.     The characteristic depth of both catalogues is $\sim33$ $h^{-1}$Mpc \citep{WatFelHud2009, FelWatHud2010, AgaFelWat2012}.  The CF2 catalogues were assembled by a compilation of Type Ia Supernovae (SNIa) \citep{TonSchBar}, Spiral Galaxy Clusters (SC) TF clusters \citep[]{GioHaySal1998, DalGioHay1999}, Streaming Motions of Abell Clusters (SMAC) FP clusters \citep{HudSmiLuc1999, HudSmiLuc2004}, Early-type Far Galaxies (EFAR) FP clusters \citep{ColSagBur2001}, TF clusters \citep{Willick1999}, the SFI++ catalogue \citep{MasSprHay2006, SprMasHay2007, SprMasHay2009}, group SFI++ catalogue \citep{SprMasHay2009}, Early-type Nearby Galaxies (ENEAR) survey \citep{daCBerAlo2000, BerAlodaC2002, WegBerWil2003} and a surface brightness fluctuations (SBF) survey \citep{TonDreBla2001}. 

The CF3-galaxy catalogue \citep{TulCouSor2016} contains 17669 galaxies, including all the CF2 galaxy distances together with 2257 distances derived from the correlation between galaxy rotation and luminosity with photometry at 3.6$\mu m$ obtained with Spitzer Space Telescope and 8885 distances based on the Fundamental Plane sample derived from the Six Degree Field Galaxy Survey (6dFGS) \citep{SprMagCol2014}. The CF3-group catalogue contains 11878 groups and galaxies.

We select the galaxies for each mock survey by ensuring a best fit to the radial selection function of the real catalogue, parametrized as \citep[see also][]{FKP}
\begin{equation}
\label{eq:selection_function}
f(r) = \mathscr{A}\left(\frac{r}{r_0}\right)^{n_1}\left[1+\left(\frac{r}{r_0}\right)^{n_1+n_2}\right]^{-1}\, ,
\end{equation}
where $\mathscr{A}$ is the scaling factor, which depends on the number of galaxies and the bin size. The constants $n_1$ and $n_2$ are the powers that fit the volume- and magnitude-limited regions, respectively, and $r_0$ is the distance at which the number of galaxies is maximum. At $r<(>)\ r_0$, volume (magnitude)--limited effects dominate, respectively.

\begin{figure}
\centering
\includegraphics[scale=0.41]{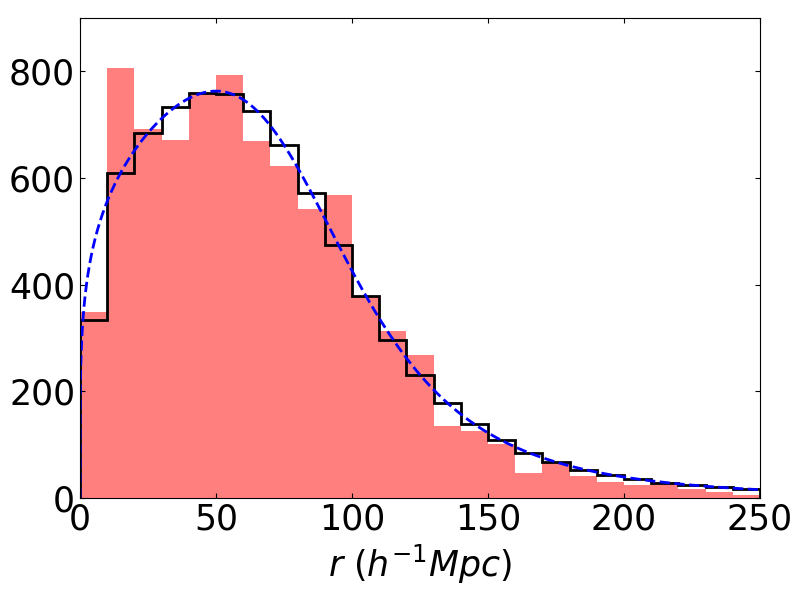}
\caption{The radial distribution of the CF2-galaxy survey (red histogram) and an example of one of its mock catalogues (solid line histogram). The dashed blue line indicates the selection function for CF2-galaxy survey given in Eq.~\ref{eq:selection_function} with parameter values $\mathcal{A}=933$, $r_0=96$ $h^{-1}$Mpc, $n_1=0.23$, and $n_2=4.25$. The bin width is 10$h^{-1}$Mpc.  For the real survey, the fit has a $\chi^2=191$ for 25 degrees of freedom, whereas for the mock survey it is $\chi^2=41.7$.}
\label{fig:radialdist}
\end{figure}

The galaxies in the real surveys are binned by their estimated distance, and the selection function given in Eq.~\ref{eq:selection_function} is fit to the radial distribution by the least-squares method, as shown in Fig.~\ref{fig:radialdist} (the smooth dashed curve). The resulting fit is then used to select the correct number of galaxies from the Millennium Simulation at each distance. The relevant selection function parameters for each of the four catalogues we use are fairly similar. For each real survey catalogue, we generated 100 mock catalogues by placing the centres of the mocks randomly in the simulation box, which can be regarded as Copernican observers. 

We studied in detail the construction of mock catalogues, both their distribution in the simulation boxes and the parameters used to produce them. A recent study choosing local group observers was introduced by \citet{HelNusFei2017}, which provides a detailed study of cosmic variance by considering differences in velocity observables as measured by a Copernican observer and its local group (LG)--equivalents.  As discussed in Sec.~\ref{sec:CV}, the differences between the velocity correlation functions in our mock catalogs are cosmic variance dominated; the large variations between random observers thus overwhelm the difference between choosing random or LG--equivalent locations. The angular distribution of the CosmicFlows surveys is not considered in the mock catalogues, since its effect was found to be negligible on both velocity correlation statistics and cosmological parameter estimation.

The mean distance between the centres of mocks from Millennium Simulation is about $250$ $h^{-1}$Mpc, which is not large enough for the mocks to be completely independent. However, we also tested mocks from the MultiDark Planck 2 (MDPL2\footnote{https://www.cosmosim.org/cms/simulations/mdpl2/}) Simulation (box size equals 1000 $h^{-1}$Mpc, mean centre separation about $680$ $h^{-1}$Mpc) and found no significant differences in the results.

\section{Velocity Correlations}
\label{sec:correlation_function}

Measuring the velocity correlation tensor  $\Psi_{ij}(r) = \left<v_i(r_1)v_j(r_2)\right>$ directly is untenable in practice, since we can only measure the radial component of a galaxy velocity.   \cite{GorDavStr1989} got around this problem by introducing two velocity correlation statistics that use only the radial peculiar velocity, $u= {\bf v}\cdot {\bf \hat r}$, 
\begin{eqnarray}
\label{eq:psi1}
\psi_1(r)  &\equiv& \frac{\Sigma_{{\rm pairs}({r})}u_1u_2\cos\theta_{12}}{\Sigma_{{\rm pairs}({r})}\cos^2\theta_{12}}\, ,\\
\label{eq:psi2}
\psi_2 (r) &\equiv& \frac{\Sigma_{{\rm pairs}({r})}u_1u_2\cos\theta_{1}\cos\theta_2}{\Sigma_{{\rm pairs}({r})}\cos\theta_{12}\cos\theta_1\cos\theta_2}\, .
\end{eqnarray}
where $r=|\vect{r_2}- \vect{r_1}|$ is the scalar separation between galaxies and the sums are performed over separation bins, the quantities $u_1$ and $u_2$ are the radial peculiar velocities of the first and second galaxy in a given pair, respectively.  The angles $\theta_1$ and $\theta_2$ are the angles between the position vectors of galaxies, ${\bf r_i}$, and the vector separating them, \eg\ $\cos\theta_1 = {\bf \hat r_1}\cdot{\bf \hat r}$, where ${\bf \hat r} =\frac{{\bf r_2}- {\bf r_1}}{r}$.    The angle $\theta_{12}$ is the angle between the position vectors of the two galaxies, so that $\cos\theta_{12}= {\bf \hat r_1}\cdot {\bf \hat r_2}$.   Thus, the numerator of $\psi_1$ is the sum over the dot products of the radial peculiar velocities, while that of $\psi_2$ is the sum over the products of the components of the radial velocities along the separation vector.   

Theoretically, the velocity field should be a potential flow field \citep{BerDek1989}, so that the radial peculiar velocity field should in principle carry all of the information contained in the full 3-D velocity field.  Indeed, \cite{GorDavStr1989} showed that when making this assumption, the radial and transverse 
velocity correlation tensor $\Psi_{ij}=\langle v_iv_j\rangle$ can be recovered from the statistics $\psi_1$ and $\psi_2$.   
We begin by expressing the correlation between radial peculiar velocities in terms of the velocity correlation tensor,
\begin{equation}
\langle u_{1} u_{2}\rangle = \hat r_{1}\hat r_{2} \langle v_{1i} v_{2j}\rangle \hat r_{1i}\hat r_{2j}\, ,
\label{eq:uu}
\end{equation}
\cite{Gorski1988} showed that the full velocity correlation tensor can be written in terms of two independent functions $\Psi_\parallel (r)$ and $\Psi_\perp (r)$, the correlation of the components of the velocity along and perpendicular to the vector separating two galaxies respectively,
\begin{equation}
\langle v_{1i} v_{2j}\rangle=
 \left[ \Psi_{\parallel}(r) - \Psi_\perp (r)\right] \hat r_{1i} \hat r_{2j} + \Psi_\perp (r)\delta_{ij}\, ,
\label{eq:vv}
\end{equation}
and $\vect{r}= \vect{r_2}- \vect{r_1}$.

Plugging Eqs. (\ref{eq:uu}) and (\ref{eq:vv}) into Eqs. (\ref{eq:psi1}) and (\ref{eq:psi2}) results in
\begin{eqnarray}
\label{eq:psi1_rt}
\psi_1(r) &=& A(r)\Psi_\parallel + \left[1-A(r)\right]\Psi_\perp\, ,\\
\label{eq:psi2_rt}
\psi_2(r) &=& B(r)\Psi_\parallel + \left[1-B(r)\right]\Psi_\perp\, ,
\end{eqnarray}
where functions $A(r)$ and $B(r)$ are given by
\begin{eqnarray}
\label{eq:A}
A(r)  &=& \frac{\Sigma_{{\rm pairs}({r})}\cos\theta_1\cos\theta_2\cos\theta_{12}}{\Sigma_{{\rm pairs}({r})}\cos^2\theta_{12}}\,,\\
\label{eq:B}
B(r)  &=& \frac{\Sigma_{{\rm pairs}({r})}\cos^2\theta_1\cos^2\theta_2}{\Sigma_{{\rm pairs}({r})}\cos\theta_{12}\cos\theta_1\cos\theta_2}\,.
\end{eqnarray}
Note that the functions $A$ and $B$ are independent of the velocities and only depend on the distribution of the galaxies in the sample.

Inverting these relations, we can express the parallel and perpendicular components of the velocity correlation as simple linear combination of $\psi_1$ and $\psi_2$, 
\begin{eqnarray}
\label{eq:PSI_r}
\psi_{\parallel}(r) &=& \frac{\left[1-B(r)\right]\psi_1(r) - \left[1-A(r)\right]\psi_2(r)}{A(r) - B(r)}\, ,\\
\label{eq:PSI_t}
\psi_{\perp}(r) &=& \frac{B(r)\psi_1(r) - A(r)\psi_2(r)}{B(r)-A(r)}\, .
\end{eqnarray}
These relations allow us to estimate the physically meaningful 3-D velocity correlations from catalogues using only measurements of radial peculiar velocities.  

Eqs.~(\ref{eq:psi1}), ({\ref{eq:psi2}), (\ref{eq:A}), (\ref{eq:B}), (\ref{eq:PSI_r}) and (\ref{eq:PSI_t}) all involve sums over pairs of galaxies whose separation falls within a given bin.   Thus it is important to determine the positions of galaxies as accurately as possible.   Here, we will follow previous researchers \citep{GorDavStr1989,BorCosZeh2000} and use redshift to estimate galaxy distances and hence galaxy separations. Redshift provides more accurate distances than distance indicators for all but the closest galaxies in our catalogues and also removes the need for Malmquist bias corrections \citep[e.g.][]{DavNusWil1996, WilStrDek1997,TulCouSor2016}.   

\section{Variance Analysis}
\label{sec:error}

To use velocity correlation statistics effectively, one needs to know how much they can be expected to vary between different locations and from sample to sample due to measurement errors.   While it would be ideal to estimate variances theoretically,  the sums over pairs form of the statistics make this extremely difficult. Here we will estimate the variances using mock surveys from the Millennium Simulation data. As mentioned before, we also used the MDPL2 simulation to verify that the results are robust. The two main components of the variance of the velocity correlation statistics are the cosmic variance, which can be estimated using mock surveys drawn from different locations in the simulation box, and measurement uncertainty, which can be estimated by making multiple realizations of a single survey with randomly generated errors. 

\subsection{Cosmic Variance}
\label{sec:CV}

In section~\ref{sec:catalogue}, we discussed generating one hundred simulation catalogues for each real survey.   Since the simulation box is much larger than the survey volume, if we draw mock surveys centred on random galaxies in the box we can study how much our correlation statistics are affected by cosmic variance.  

In Fig.~\ref{fig:CF2-gal-CV}, we show the mean and variance of the velocity correlation function with 500 km s$^{-1}$ bin width calculated from CF2-galaxy mock catalogues.  In order to match our calculations with real catalogues, we have calculated the correlation function using the redshift instead of distance as discussed above.  Using CF2-groups instead of CF2-galaxies makes very little difference in our results.   

Comparing the six plots in Fig~\ref{fig:CF2-gal-CV}, we might be surprised that the velocity-independent functions $A(r)$ and $B(r)$ also show cosmic variance.  This is due to the fact that even though the simulation catalogues have the same radial selection functions, each mock catalogue has a slightly different distribution of galaxies due to the particular locations of the galaxies in the region of the simulation box where the catalogue was drawn from.  

\begin{figure}
\centering
\includegraphics[scale=0.28]{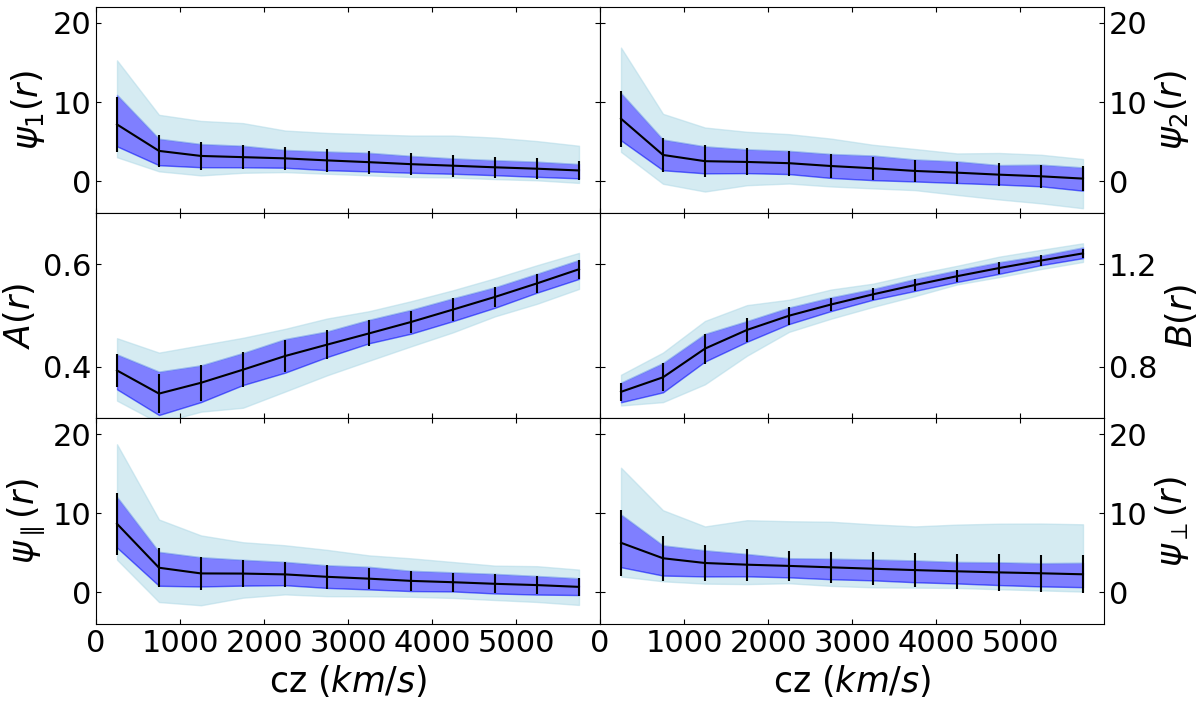}
\caption{\label{fig:CF2-gal-CV} The mean and cosmic variance of $\psi_1$, $\psi_2$, $A(r)$, $B(r)$, $\psi_{\parallel}$ and $\psi_{\perp}$, of 100 mock catalogues with CF2-galaxy distribution. $\psi_1$, $\psi_2$, $\psi_{\parallel}$ and $\psi_{\perp}$ are in units of (100 km s$^{-1})^2$. The centre solid line and error bars show the average and standard deviation of each function. The contours indicate the regions containing $68\%$ and $95\%$ of the results. The bins in this plot are 500 km s$^{-1}$ wide. }  
\end{figure}

Several things are notable in Fig.~\ref{fig:CF2-gal-CV}.   First, we see that the statistic $\psi_2$ appears to be well behaved on CosmicFlow type surveys, $\psi_2$ of CF3 shows similar result. This is in contrast to the \citet{GorDavStr1989} findings (fig. 2 in \cite{GorDavStr1989}) , who neglected to use this statistic since it was deemed less stable when applied to the catalogues that were available at the time.   

Further, we see that the cosmic variance in the velocity correlation function is large, of the order of the mean, and the distribution of cosmic variance is skewed, and thus non-Gaussian. This can be seen most clearly from the 95\% contours in Fig.~\ref{fig:CF2-gal-CV}, which are not symmetric about the mean.   We can understand the skewness by noticing that the correlation function is quadratic in the velocity, which is itself a Gaussian distributed variable. Thus we expect the cosmic variance of the correlation functions to have  a bivariate Gaussian distribution, that is, a Wishart distribution \citep{Wishart}, which is a generalization of a $\chi^2$-distribution.  Like the $\chi^2$ distribution, this distribution has exponential tails which fall off much more slowly than Gaussian tails. We expect that the correlation functions also contain contributions from noise, which we would expect to be Gaussian distributed.

\begin{figure}
\centering
\includegraphics[scale=0.5]{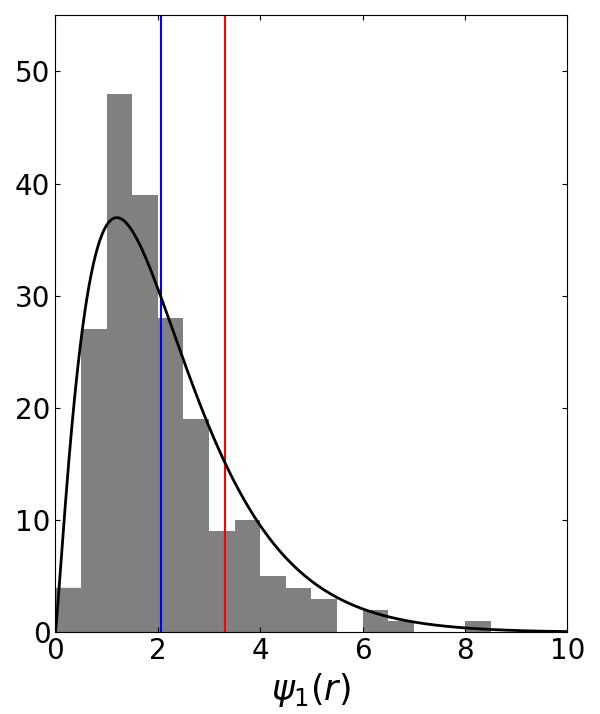}
\caption{\label{fig:CV-distribution} The distribution of $\psi_1$ in 4000-4500 km s$^{-1}$ bin of 200 mock catalogues, in units of (100 km s$^{-1})^2$. The blue vertical line on the left is the mean of the mock catalogues ($\overline{\psi_1}$); the red vertical line on the right is the CF3-galaxy catalogue measurement in the bin; the black solid curve is the Wishart distribution fitting.}
\end{figure}

In general, then, the variance in the correlation function has a distribution that is a convolution of a Wishart and a Gaussian, as the example in Fig.~\ref{fig:CV-distribution} clearly shows. The skewness becomes more non-Gaussian in larger separation bins, due to the lower number of degrees of freedom, in that fewer modes are contributing to the correlation function. The non-Gaussian distribution of the cosmic variance,  in particular, the exponential tails of the distribution, limits the information we get using the velocity correlation function. Thus, both the size of the cosmic variance and its distribution pose challenges for using velocity correlation statistics to constrain cosmological models.

We compare the cosmic variance of mocks from the Millennium and the MDPL2 Simulations, with and without including an angular mask (\ie\ reproducing the CF3 angular distribution). We find the mean measurement and the contours of the cosmic variance are all within the uncertainties and show little differences across the samples. The cosmic variance from those four kinds of mocks are combinations of Wishart and Gaussian distributions; however, their distribution parameters may be different, which leads to variations in the cosmological constraints. The cosmic variance distribution is somewhat sensitive to non-Gaussian skewness (\eg\ bin width, correlation truncations, and mock selections), which leads to some differences in cosmological parameter constrains. This topic will be discussed further in section~\ref{sec:linear}.

\begin{figure}
\centering
\includegraphics[scale=0.33]{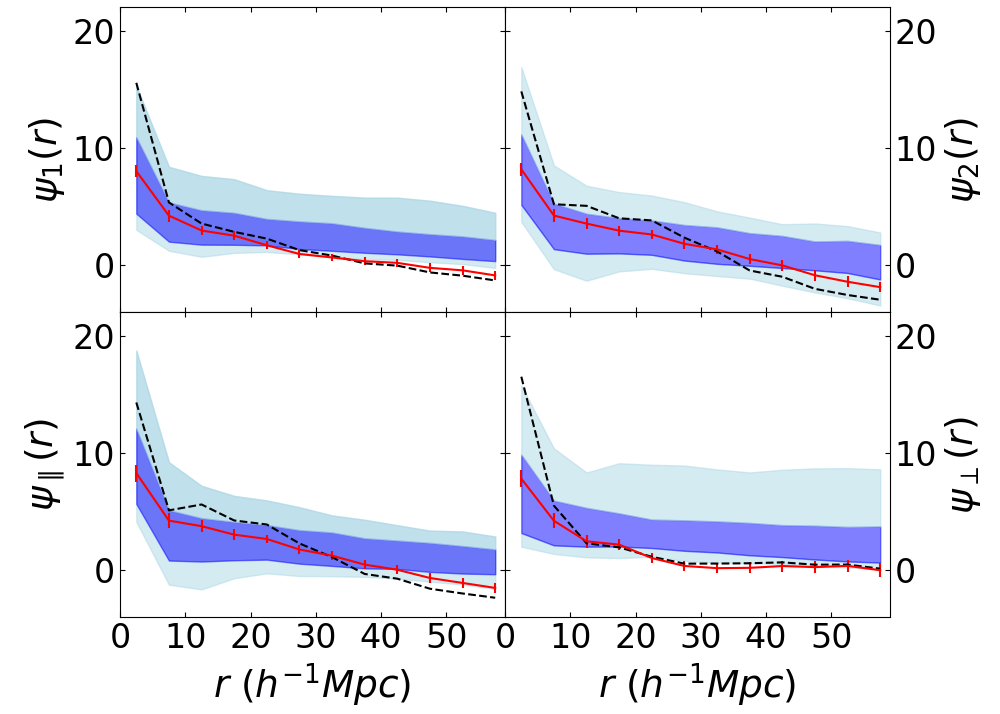}
\caption{\label{fig:CF2-gal-se-m-ve} The result of perturbing the distance moduli of one CF2-galaxy simulation catalogue, in units of (100 km s$^{-1})^2$. The bin width is 500 km s$^{-1}$ (redshift). The black dashed line shows the original simulation catalogue; the red solid line with error bars is the average of the perturbed catalogues. The error bars show the standard deviation of the perturbed results, which is regarded as the measurement error. The background contours indicate the cosmic variance from Fig.~\ref{fig:CF2-gal-CV}.}
\end{figure}

\subsection{Measurement Error}
\label{sec:ME} 

The other main source of uncertainty in the correlation functions comes from the uncertainty of the peculiar velocity measurements. Because the peculiar velocity measurements from the simulation do not have uncertainty, we used a Monte Carlo method to simulate distance measurement errors, which are also manifested as uncertainties in radial peculiar velocities.  We generate 100 error analysis catalogues of each mock with distance perturbed by a constant Gaussian variance, which is about 20\% \citep[e.g.][]{MasSprHay2006, SprMasHay2007, TulCouDol2013}; however, this uncertainty is non--Gaussian, since it originates from the uncertainty of the distance modulus. Thus, we simulate the measurement error by generating a set of 100 versions of each mock catalogue with distance moduli perturbed with a constant Gaussian variance $\sigma=0.43$, which roughly corresponds to $\sim$20\% distance error. The perturbed distances are used to calculate new velocities using the unbiased method described in \citet{WatFel2015} so that the errors in the velocities are Gaussian distributed.  The uncertainties in the correlation function due to measurement errors are estimated by averaging the standard deviations of the correlations calculated from the perturbed data sets.   In Fig.~\ref{fig:CF2-gal-se-m-ve}, we show the results of perturbing a single CF2-galaxy catalogue in this way.  From the figure it is evident that measurement errors affect our results much less than cosmic variance (see also Fig.~\ref{fig:size_sample_err}).   

\begin{figure}
\centering
\includegraphics[scale=0.5]{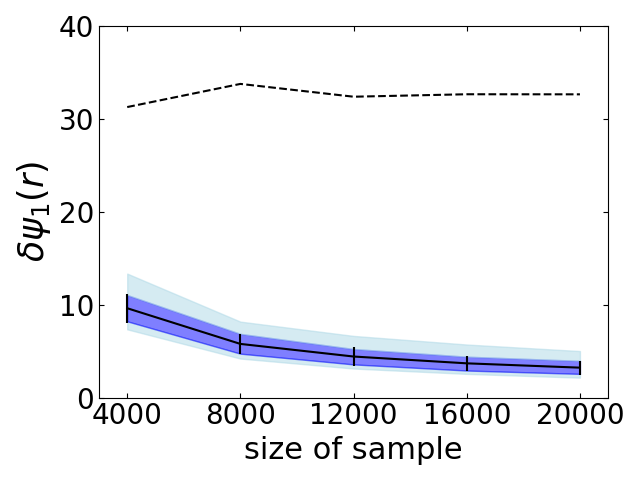}
\caption{\label{fig:size_sample_err} The dashed line shows the standard deviation of the cosmic variance of the correlation statistic $\psi_1$ at the redshift separation bin 0 - 500  km s$^{-1}$ of the simulation catalogues as a function of sample size.   The solid line indicates measurement error of the perturbed sample averaged over all the bins as a function of sample size, in units of 1000 km$^2$ s$^{-2}$. The contours show the 68\% and 95\% ranges of the results, whereas the error bars are the standard deviations of the measurement errors.}
\end{figure}

\subsection{Measurement Error and sample size}
\label{sec:ESS}

Once we have a large enough sample size to adequately probe a volume, we do not expect cosmic variance to change with sample size, since in this case it is due to the variations between volumes of a similar size.   To decrease the cosmic variance we would need to increase the depth of our sample; larger volumes should vary less as we approach the scale of homogeneity for the Universe.   

In contrast, we can reduce the effect of measurement errors by increasing our sample size, even if the volume being probed stays the same.  To characterize the effect of sample size on measurement errors, we select new mock surveys with up to $20,000$ galaxies by changing $\mathcal{A}$ in Eq.~(\ref{eq:selection_function}) while leaving the other selection function parameters unchanged.   We then perturb them using the same method  described in section~\ref{sec:ME}.   In Fig.~\ref{fig:size_sample_err} we show how the cosmic variance and measurement errors of the correlation statistic $\psi_1$ changes with sample size. Increasing the sample size reduces the statistical errors significantly, while, as expected, the cosmic variance doesn't show significant difference.   

\section{Correlation Results}
\label{sec:CR}

Fig.~\ref{fig:size_sample_err} shows the relationship between sample size and statistical error. Using this relation, we can calculate the statistical error of CF3 surveys, which have many more data points than the CF2 surveys. Fig.~\ref{fig:CF3-gal-h750} shows the correlation function of CF3-galaxy survey with Hubble constant $H_0=$ 75 km s$^{-1}$ Mpc$^{-1}$; this is the value given by \cite{TulCouSor2016} as the best fit to the data. We see that the correlation function of CF3 is larger than the mean value of correlations of mock catalogues. \citet{HelNusFei2017} show that cosmic variance of local group (LG) observer mocks with a Virgo like cluster is larger than the Copernican observer mocks without Virgo, which explains the smaller mean correlation value from mocks in Fig.~\ref{fig:CF3-gal-h750}.

\begin{figure}
\centering
\includegraphics[scale=0.29]{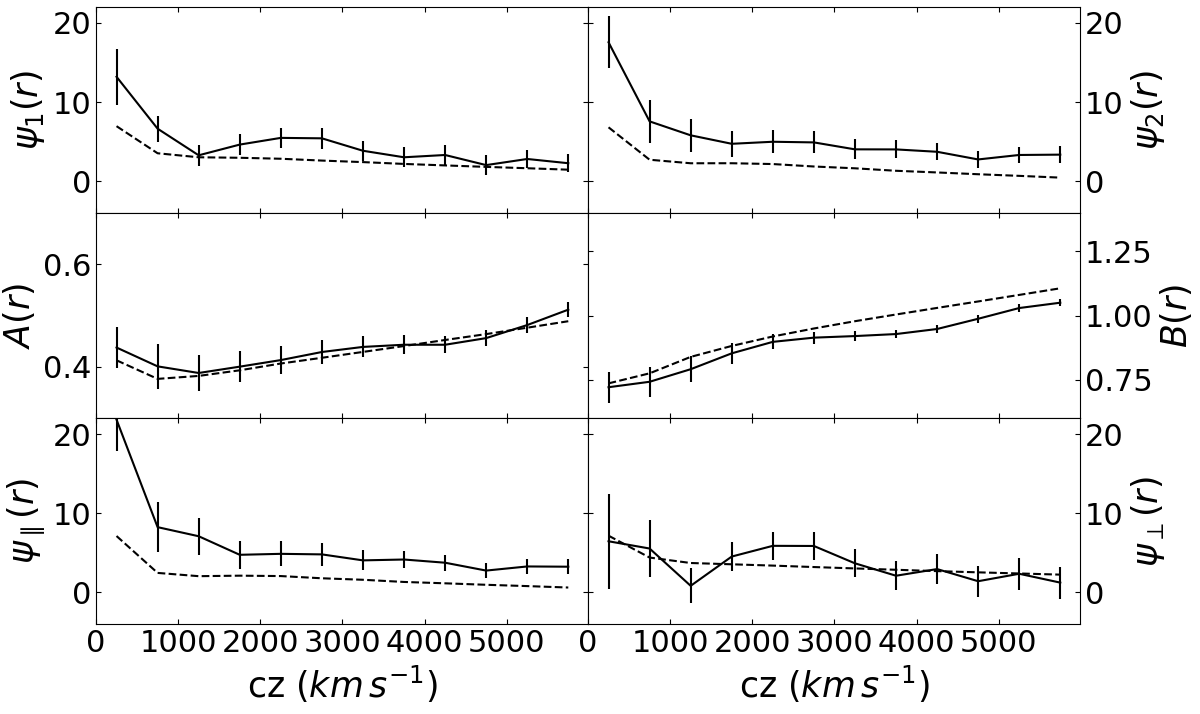}
\caption{\label{fig:CF3-gal-h750} Velocity correlation functions of CF3-galaxy survey with $H=75$km s$^{-1}$ Mpc$^{-1}$. $\psi_1$, $\psi_2$, $\psi_{\parallel}$ and $\psi_{\perp}$ are in units of (100 km s$^{-1})^2$. The solid line shows the real survey result, the error bar indicates its uncertainty, combining the effects of cosmic variance and measurement error. The dashed line indicates the average of 100 mock catalogues.}
\end{figure}

The  $75$ km s$^{-1}$ Mpc$^{-1}$ value for the Hubble constant is in significant tension with the larger-scale \cite{Planckparameters2014} result of $67.74\pm 0.46$ km s$^{-1}$ Mpc$^{-1}$ obtained from the cosmic background radiation.  One possible source of this discrepency is an error in the calibration of the distance indicators used in the CF3 measurement.   Given that distance indicators are calibrated in sequence working outwards, this would most likely be an error near the base of the distance ladder; for example, with the calibration of cepheid distances \citep[see \eg][for a discussion of this tension]{RieMacHof2016}.   However, a second possibility is that our local volume has a significant outflow which is artificially  inflating the value of $H_0$.   \cite{TulCouSor2016} discuss this possibility qualitatively by assessing the size and radial profile of the outflow required to obtain different underlying values of $H_0$.  
Given that radial flows also impact the correlation function, we can preform a similar analysis here by varying the Hubble constant used in estimating velocities from distances in the CF3 survey.  

\begin{figure}
\centering
\includegraphics[scale=0.27]{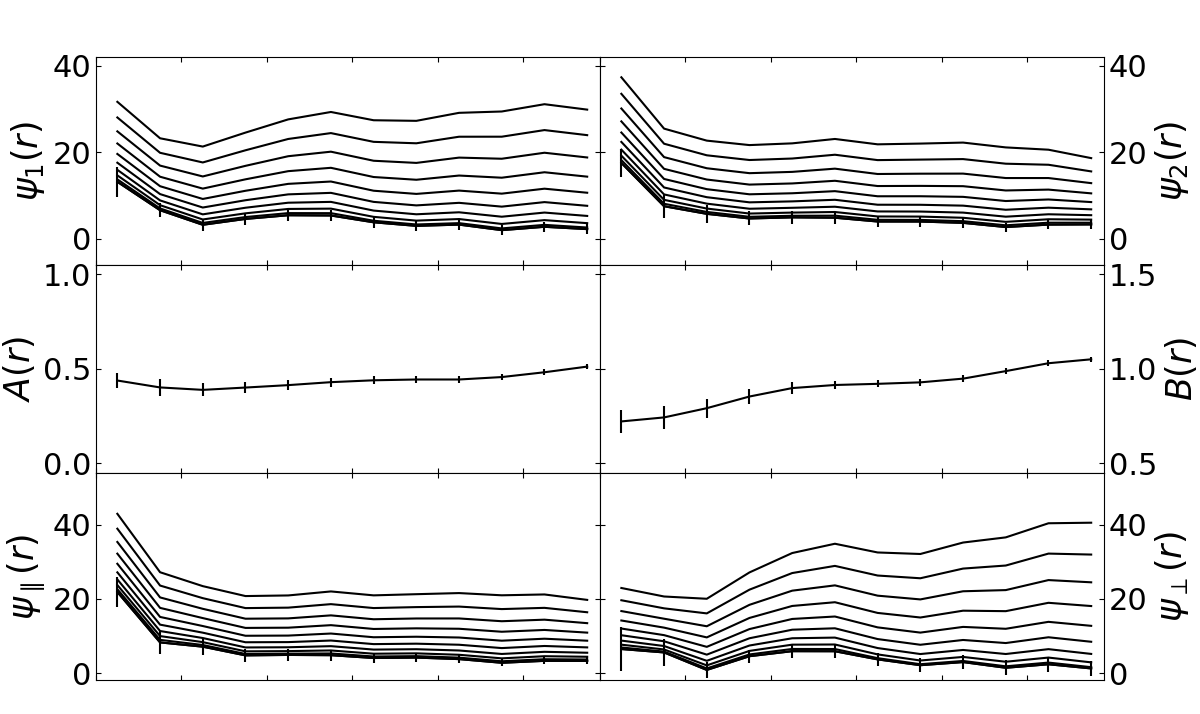}
\includegraphics[scale=0.27]{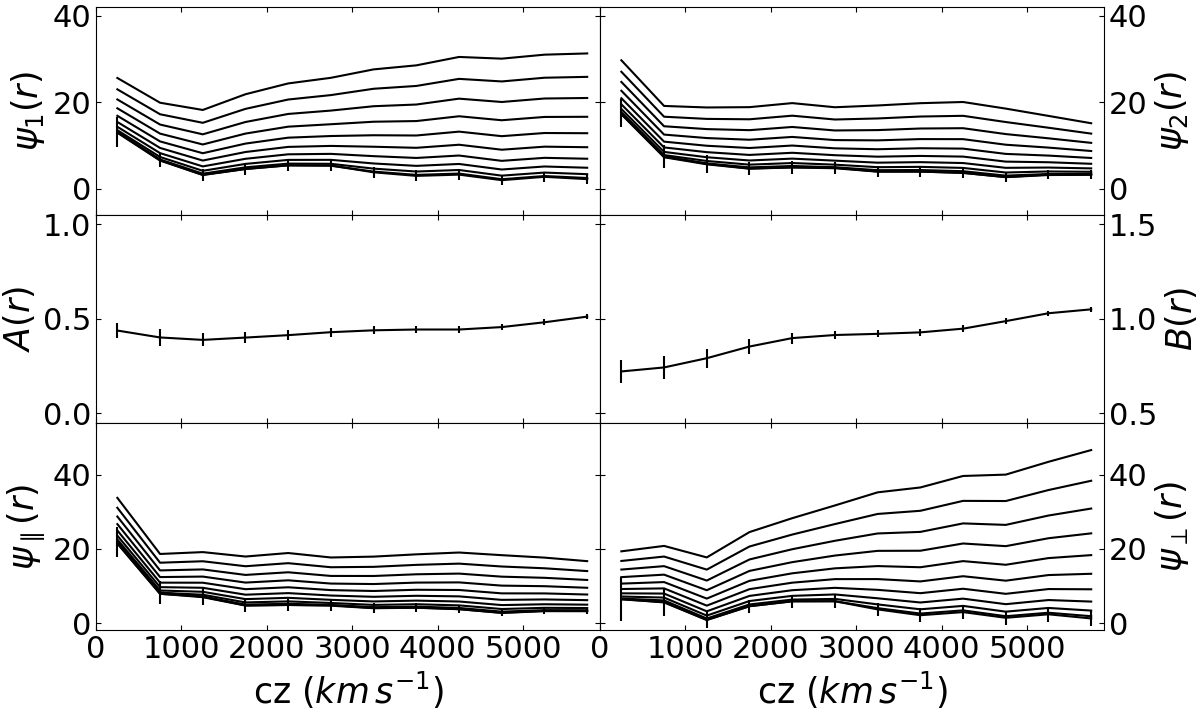}
\caption{\label{fig:CF3-gal-7075} {\bf Top Panel:} Velocity correlation functions of CF3-galaxy surveys with Hubble constant from 70 to 75 km s$^{-1}$ Mpc$^{-1}$ (top to bottom, 0.5 km s$^{-1}$ Mpc$^{-1}$ each). $\psi_1$, $\psi_2$, $\psi_{\parallel}$ and $\psi_{\perp}$ are in units of (100 km s$^{-1})^2$. The contours indicate the cosmic variance.  {\bf Bottom Panel:} Same as the top panel, but Hubble constant from 75 to 80 km s$^{-1}$ Mpc$^{-1}$ (bottom to top, 0.5 km s$^{-1}$ Mpc$^{-1}$ each).}
\end{figure}

In Fig.~\ref{fig:CF3-gal-7075} we show the correlation functions assuming $H_0$ values from 70-75 km s$^{-1}$ Mpc$^{-1}$ in the top panel and 75-80 km s$^{-1}$ Mpc$^{-1}$ in the bottom panel.  As can be seen, even small deviations from the value of 75 km s$^{-1}$ Mpc$^{-1}$ can lead to unrealistically large correlations on a wide range of scales.   Particularly troubling is the increase in $\psi_1$ (or equivalently $\psi_\perp$) with increasing scale.   This analysis confirms the conclusion of \cite{TulCouSor2016} that it is unlikely that outflows can be the source of the discrepancy in the value of the Hubble constant between the Planck result and more local probes.      

\section{Linear Theory}
\label{sec:linear}

In this section, we explore the use of the correlation function to constrain cosmological models. This was previously attempted by \cite{BorCosZeh2000}; however, they incorrectly assumed that the cosmic variance in the correlation function $\psi_1$ was Gaussian distributed. We examine the implications that the cosmic variance in $\psi_1$ is, to a good approximation, Wishart distributed, as discussed in section ~\ref{sec:CV}.  

\begin{figure}
\centering
\includegraphics[scale=0.5]{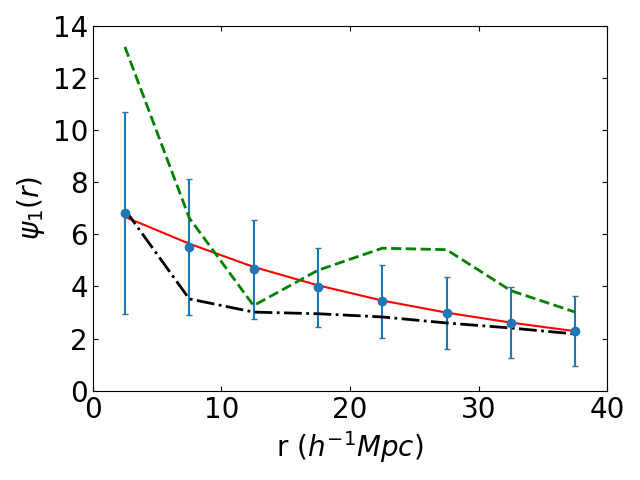}
\caption{\label{fig:linear_simual} $\psi_1$ values  of CF3-galaxy catalogue, mock catalogues and linear theory predictions, in units of (100 km s$^{-1})^2$. The blue dots show the mean of 100 mock catalogues using distance separation, the black dash dotted line shows the mean of the mock catalogues using redshift separation, the red solid line indicates the linear theory prediction, and the green dashed line shows the CF3-galaxy correlation value using redshift separation. The blue error bars represent the cosmic variance.}
\end{figure}

Linear theory describes the relation between the radial and transverse correlation functions and the power spectrum of density fluctuation $P(k)$ \citep{BorCosZeh2000}. The radial and transverse velocity correlations can be written in terms of the power spectrum as follows:

\begin{eqnarray}
\label{eq:psi_para}
\psi_\parallel(r)  &=& \frac{\beta ^2 H_0^2}{2\pi^2} \int P(k)\left[j_0(kr) - 2\frac{j_1(kr)}{kr}\right]dk\, ,\\
\label{eq:psi_perp}
\psi_\bot(r) &=& \frac{\beta ^2 H_0^2}{2\pi^2} \int P(k)\frac{j_1(kr)}{kr} dk ,
\end{eqnarray}
where $j_i(kr)$ is the i$^{\rm th}$-order spherical Bessel function, $P(k)$ is the power spectrum, and on large scales $\beta = f(\Omega_m)\sigma_8$ \citep{GuzPieMen2008,Turnbull2012}, $f(\Omega_m)=\Omega_m^{0.55}$ \citep{Linder2005} and $\sigma_8$, is the amplitude of density fluctuations on a scale of $8 h^{-1}$Mpc.   For $P(k)$ we use the parametrization of \cite{EisHu1998} that expresses $P(k)$ in terms of the matter density $\Omega_m$, the baryon density $\Omega_b$, and the Hubble parameter $h$.  We normalize the power spectrum using the value of $\sigma_8$.   We checked that $P(k)$ obtained from this parametrization is in good agreement with those produced by more sophisticated methods, such as \citet{HeiWhiWagHab2010,HeiLawKwaHab2014,AgaAbdFel2012,AgaAbdFel2014}.      

\begin{figure}
\centering
\includegraphics[scale=0.34]{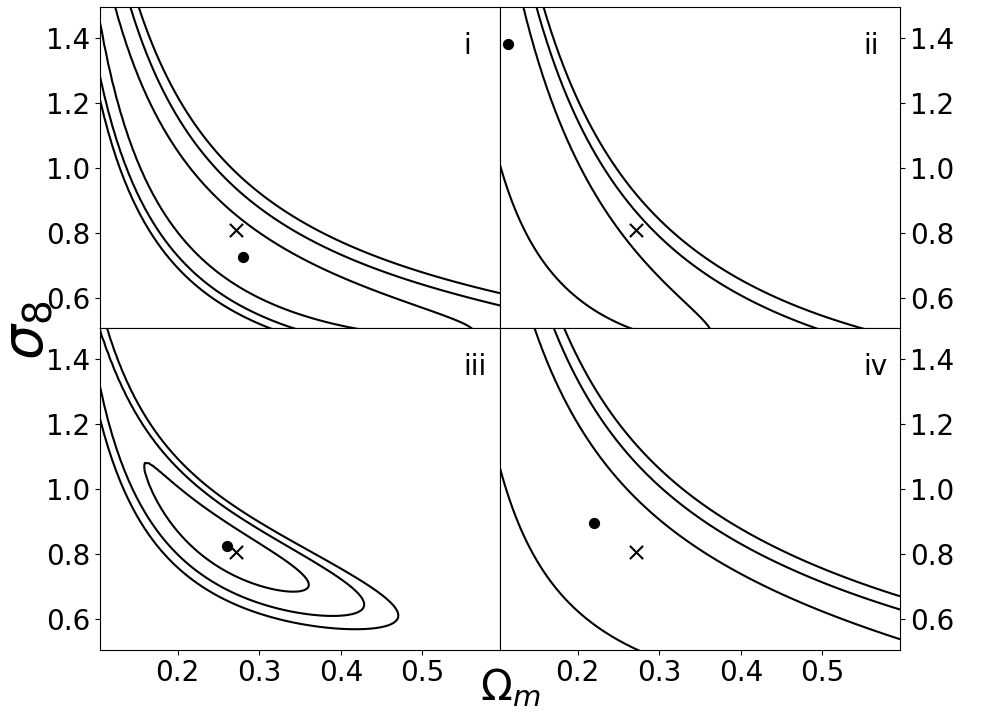}
\caption{\label{fig:chi_multi} $\chi^2$ plots of four weighting schemes of mock catalogue correlations in redshift and linear theory predictions truncated at $4000$ km s$^{-1}$ with $500$ km s$^{-1}$ binwidth. The minimum $\chi^2$ value has been subtracted from each cell. The dot  indicates the best $\chi^2$ fitting, and the contours show 68\% and 95\% and 99.9\%  likelihood of $\chi^2$ values. The cross indicates the value of the Millennium simulation (Table~\ref{T_Mill}).}
\end{figure}

Linear theory reproduces the average correlation functions from the mock catalogues quite well if true distances are used to specify positions rather than redshift. In Fig~\ref{fig:linear_simual}, we show $\psi_1$ obtained from linear theory using the simulation cosmological parameters from Table~\ref{T_Mill} together with the mean $\overline{\psi_1}$ from the mock catalogues.   We used the average $A$ and $B$ functions from the mock catalogues to go from the linear theory $\psi_\parallel$ and $\psi_\perp$ to $\psi_1$ via Eq.~\ref{eq:psi1_rt}.   We see that linear theory prediction matches the averages of the mock catalogues well when using distance seperation. However, when redshift is used to specify distance, we see the effects of redshift distortion, although even in this case the effects are smaller than the cosmic variance. Redshift distortions also affect the estimation of cosmological parameters and we discuss their inclusion below.

The velocity correlation functions measured using the CF3 catalogues are within the cosmic variance of those using mock catalogues from the simulations, which in turn use initial conditions close to those parameters measured by seven-year Wilkinson Microwave Anisotropy Probe \citep[WMAP][see also Table~\ref{T_Mill}]{WMAP7}.   Thus, the CF3 correlation function appears to agree with the cosmological standard model.  We can make this assessment more quantitative by developing a $\chi^2$ statistic for the difference between the measured correlation function and the prediction from linear theory.   This analysis results in constraints on $\Omega_m$ and $\sigma_8$, which are the main factors that determine the shape and amplitude of the power spectrum, respectively. In addition, $\chi^2$ analysis is complicated by the fact that the values of $\psi_1$ in different bins are strongly correlated, since the large-scale velocity modes lead to correlations that contribute similarly to all separations. In order to account for this correlation, we use the weighted $\chi^2$ fitting method first introduced by \citet{Kaiser1989}: 
\begin{equation}
\label{eq:chi_kaiser}
\chi^2 = \sum_{i}w(r_i)\left[ \psi^M_1(r_i) - \psi^L_1(r_i)\right]^2 ,
\end{equation}
where $w$ is the weighting function, $\psi^M_1$ is the measured value from the catalogue (CF2 or CF3), and $\psi^L_1$ is the linear prediction.

The $\chi^2$ fitting is strongly affected by the errors of velocity correlation functions:  the cosmic variance, the redshift distortion, and the measurement error, which is small enough to be neglegible. In order to explore the effects of the complicated error distributions, we choose four different weighting schemes and test them with mock catalogues: {\it i}) Linear prediction weight $w(r)=1/\psi^L_1(r)$; {\it ii}) Cosmic variance based weight using covariance matrix \citep{BorCosZeh2000}; {\it iii}) redshift distortion based weight using the redshift distortion matrix, which can be regarded as an error correlated matrix; {\it iv}) Combination of error and redshift distortion based weight using the sum of the covariance matrix and redshift distortion matrix, which leads to a full covariance matrix \citep{Blobel2003, DAgostini1995}.   

Eq.~\ref{eq:w-lin} -~\ref{eq:w-combo} below show the four weighting schemes, where $C$ is the covariance matrix; $\lambda$ is the redshift distortion matrix; $N_{mock}$ is the number of mock catalogues; $ \psi^i_{1,l}$ is the correlation value of the $i^{th}$ separation bin of the $l^{th}$ mock catalogue; $\overline{\psi}^i_1$ is the average value of $N_{mock}$ catalogues in the $i$th separation bin; $ \psi^{i,s}_{1,l}$ ($\psi^{i,r}_{1,l}$) is the correlation value of the $i^{th}$ bin of the $l^{th}$ mock catalogue using redshift (distance) separation.

\begin{strip}
\begin{eqnarray}
\label{eq:w-lin}
i).& \chi^2 &= \sum_{i}\frac{\left[ \psi^M_1(r_i) - \psi^L_1(r_i)\right]^2}{\psi^L_1(r_i)}\\
\label{eq:w-cov}
ii).& \chi^2 &= \sum_{i,j}\left[ \psi^M_1(r_i) - \psi^L_1(r_i)\right] C^{-1}_{ij}\left[ \psi^M_1(r_j) - \psi^L_1(r_j)\right] \ \ \ \ \    C_{ij} = \frac{1}{N_{mock}} \sum^{N_{mock}}_{l=1} \left( \psi^i_{1,l} - \overline{\psi}^i_1\right) \left( \psi ^j_{1,l} - \overline{\psi}^j_1\right)\\
\label{eq:w-redshift}
iii).& \chi^2 &=  \sum_{i,j}\left[ \psi^M_1(r_i) - \psi^L_1(r_i)\right] \lambda^{-1}_{ij}\left[ \psi^M_1(r_j) - \psi^L_1(r_j)\right]  \ \ \ \ \    \lambda_{ij} = \frac{1}{N_{mock}} \sum^{N_{mock}}_{l=1} \left( \psi^{i,s}_{1,l} - \psi^{i,r}_{1,l}\right) \left( \psi ^{j,s}_{1,l} - \psi^{j,r}_{1,l}\right)\\
\label{eq:w-combo}
iv).& \chi^2 &=  \sum_{i,j} \left[ \psi^M_1(r_i) - \psi^L_1(r_i)\right] \left( C_{ij} + \lambda_{ij}\right) ^{-1} \left[ \psi^M_1(r_j) - \psi^L_1(r_j)\right]
\end{eqnarray}
\end{strip}

In Fig.~\ref{fig:chi_multi}, we show the $\chi ^2$ estimates of the mock catalogue correlations and the estimates for the parameters $\Omega_m$ (x-axis) and $\sigma_8$ (y-axis) for each of the weighting schemes. (The 68\% and 95\% contours are defined by the same method used in \cite{BorCosZeh2000}.) 

\begin{figure}
\centering
\includegraphics[scale=0.34]{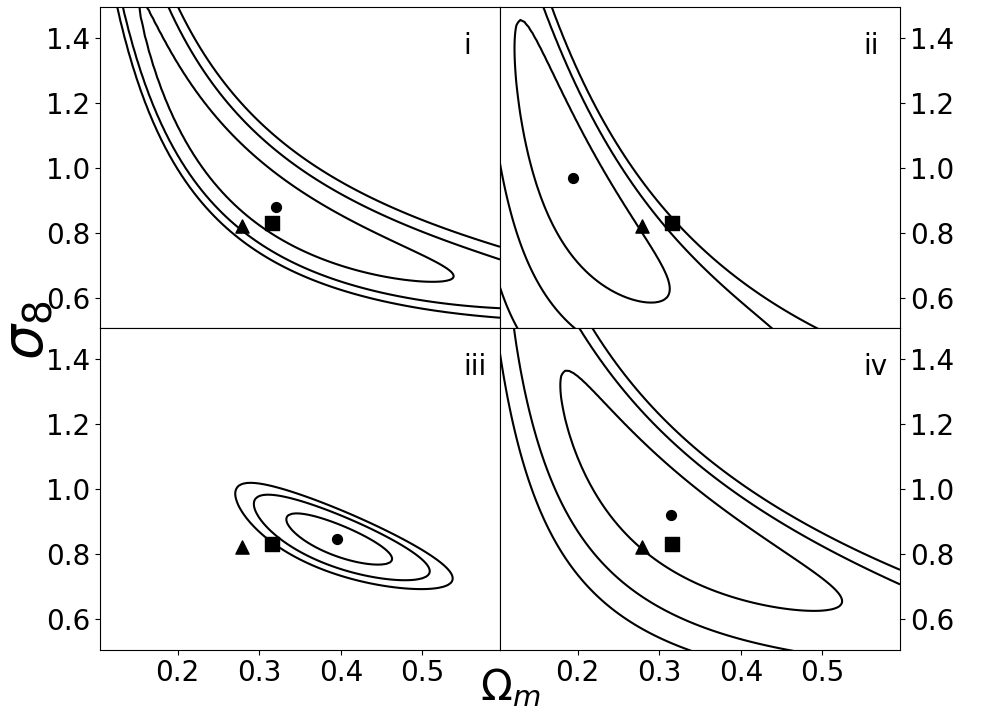}
\caption{\label{fig:chi_multi_cf3} Same as Fig.~\ref{fig:chi_multi}, but using CF3-galaxy correlations. The triangle is the WMAP result \citep{wmap9} and the square is the Planck result \citep{Planckparameters2014}.}
\end{figure}

\noindent{\it i)} Linear prediction weighted scheme (Eq.~\ref{eq:w-lin}) gives a reasonable though not very tight constraint, especially for $\Omega_m$, it is somewhat sensitive to the truncation of the correlation functions. Note that this method treats the bins as independent from each other, and thus does not incorporate the strong correlations among different bins. 

\noindent{\it ii)} Covariance matrix weighted scheme (Eq.~\ref{eq:w-cov}) does not provide strong constraints. That is because the covariance matrix is dominated by cosmic variance which is decreasing for increasing separation, and thus larger separations are given more weight in the $\chi^2$ fitting scheme. Furthermore, the non-Gaussian skewness also becomes more pronounced in large separation bins (see Sec.~\ref{sec:CV}) and thus biases the results. In addition, the Wishart error distribution is sensitive to different mock selections. Comparing the Millennium and MDLP2 mocks with and without the angular mask, we found that the simulation mocks affect the covariance matrix estimates since the covariance matrix weighted scheme is sensitively dependent on the Wishart distribution parameters for the determination of cosmic variance (see Sec.~\ref{sec:CV}). However, differences caused by the different simulation mocks do not show any trend of improving the constraints, instead, they are somewhat random. Non--Gaussian skewness is the dominant source of bias for the covariance matrix weighted scheme. 

\noindent{\it iii)} Redshift distortion weighted scheme (Eq.~\ref{eq:w-redshift}) provides a good estimate of the parameters while using mock catalogues only and the contours are more compacted than other schemes.

\noindent{\it iv)} We combine the effects of cosmic variance and redshift distortion. The constraining result of combined weighting scheme is not as tight as methods {\it i} and {\it iii}, but it does agree with the simulation parameters within 1$\sigma$.

Fig.~\ref{fig:chi_multi_cf3} shows the constraints using the CF3-galaxy.  The redshift distortion scheme by itself shows very tight contours, but disagrees with the Planck and WMAP results at a high confidence limit, which leads us to conclude that the simulation redshift distortions do not mimic the distortions in the CF3 catalogue well. In addition, the schemes become more sensitive to truncations when using real survey. Figs.~\ref{fig:truncation_O} and~\ref{fig:truncation_S} show the $\Omega_m$ and $\sigma_8$ results of the four weighting schemes with different truncations. The redshift distortion weighted scheme ({\it iii}) is more strongly dependent on the truncation choice but mostly agrees within 2 standard deviations, whereas the others agree within 1 standard deviation. 

\begin{figure}
\centering
\includegraphics[scale=0.53]{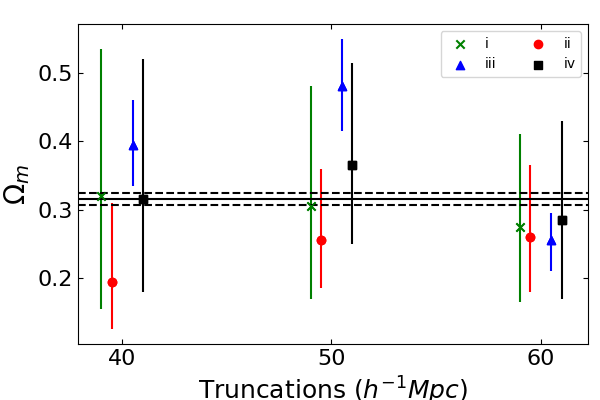}
\caption{\label{fig:truncation_O} The effect truncation has on the estimation of $\Omega_m$. The green cross markers show the constrains using the linear prediction weighted scheme ({\it i}), the red circle markers show the results using covariance matrix weighted scheme ({\it ii}), the blue triangle markers indicate the results using redshift distortion weighted scheme({\it iii}) and black square markers indicate the results using the combo weighting scheme ({\it iv}). The horizontal black lines are the $\Omega_m$ value and 1$\sigma$ determined by Planck.}
\end{figure}

\begin{figure}
\centering
\includegraphics[scale=0.53]{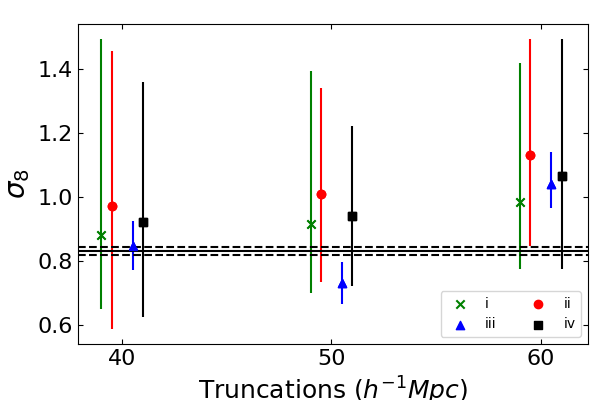}
\caption{\label{fig:truncation_S} Same as Fig.~\ref{fig:truncation_O}, but showing the $\sigma_8$ truncation results.}
\end{figure}

Considering the performances of those four weighting schemes for the simulation and observation data, we choose the scheme {\it iv} that combines the cosmic variance and redshift distortion effects in this paper to be most reliable. In Fig.~\ref{fig:chi_plot}, we show the $\chi ^2$ of CF3-galaxy correlations using the redshift distortions and cosmic variance combination scheme (Eq.~\ref{eq:w-combo}). We get $\Omega_m= 0.315^{+0.205}_{-0.135}$ and $\sigma_8=0.92^{+0.440}_{-0.295}$. As can be seen, the value of $\Omega_m$ and $\sigma_8$ agree with the results from Planck \citep[$\Omega_m=0.315\pm0.013$; $\sigma_8=0.831\pm0.013$,][]{Planckparameters2014} and WMAP9 \citep[$\Omega_m=0.279\pm0.023$; $\sigma_8=0.821\pm0.023$][]{wmap9}, whereas the value of $\sigma_8$ we get from the correlation analysis with CF3  is slightly larger but still within $1\sigma$.

Part of the discrepancy in $\sigma_8$ could be due to the fact that in the present analysis $\sigma_8$ is obtained from local data. As was clearly shown in \citet{JusFelFryJaf2010}, estimators that probe the value of $\sigma_8$ on small cosmological scales do not take into account the nonlinear evolution of the parameter at late times.  Using the parametrization from \citet{JusFelFryJaf2010}, we should compare our $\sigma_8$  results to 0.888 (Planck) and 0.879 (WMAP9) which agree to within $\sim1\sigma$.

\begin{figure}
\centering
\includegraphics[scale=0.5]{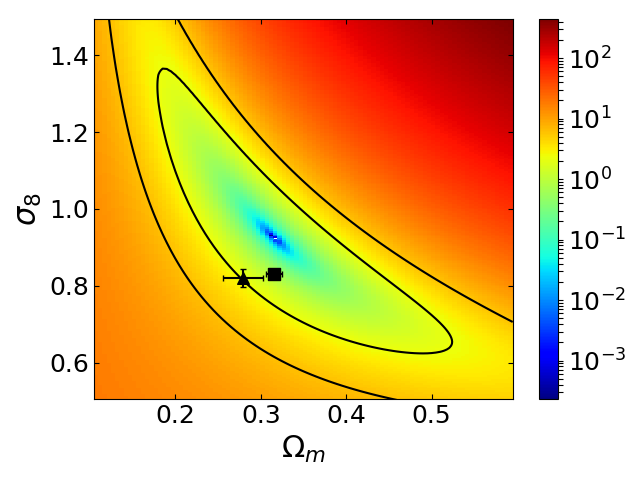}
\caption{\label{fig:chi_plot} $\chi^2$ plot for the CF3-galaxy survey with combo weighting scheme for binwidth equals $500$ km s$^{-1}$ and truncation at $4000$ km s$^{-1}$. The minimum $\chi^2$ value has been subtracted from each cell. The contours indicate 68\% and 95\% likelihood of $\chi^2$ values. The square marker indicates the best value from Planck \citep{Planckparameters2014}, whereas the triangle marker is the value of WMAP9 \citep{wmap9}. }
\end{figure}

\section{Conclusion}
\label{sec:conclusion}

In principle, the velocity Correlation Function is a powerful statistical tool in exploring the Peculiar Velocity Field and through that, the mass distribution on cosmological scales. 
We have shown that on average the correlation function calculated from simulated catalogues recovers the expected signal from linear theory, thus demonstrating that it is an unbiased statistic.   Since the statistical error in the correlation function is significantly smaller than the cosmic variance,  the velocity correlation function does a reasonable job dealing with the large uncertainty inherent in the determination of peculiar velocities of galaxies and groups.  However, the non-Gaussian nature of the cosmic variance and redshift distortion put limits on how well we can use this statistic to constrain cosmological parameters.

We have calculated the velocity correlation functions for the CosmicFlows-2 and CosmicFlows-3 catalogues and shown that they are consistent with expectations from the standard cosmological model. In addition, we have used our results together with linear theory to constrain the cosmological parameter $\Omega_m$ and $\sigma_8$. In constraining the cosmological parameters, we have assumed Gaussian distributed errors, while the simulations have clearly shown that the error distribution of the cosmic variance has distinct non-Gaussian tails. Furthermore, since the cosmic variance is smaller at larger separations, the covariance matrix gives more weight at larger separations, where skewness is most pronounced and thus,  may introduce systematic biased parameter estimations. In addition, redshift distortions give rise to the mismatch between CosmicFlows correlations and linear predictions and thus may contribute further bias to  parameter constrains. To mitigate this effect we have used a weighting scheme that combines the effects of cosmic variance and redshift distortion, which appears to be both more stable and less biased. Future studies that account for the non-Gaussian distribution of cosmic variance may result in more robust constraints, particularly with regard to uncertainties in parameter estimation. 

The systematically larger velocity correlations observed in this study, especially in closer bins, using both the CosmicFlows-2 and and  CosmicFlows-3 compilations is consistent with the observed bulk flows from these and other catalogues that is on the larger end of the expected range given the predictions from the $\Lambda$CDM model with CMB derived parameters. However, this excess may also arise from local inhomogeneities in our local volume.

\section{Acknowledgements}

The CosmoSim database used in this paper is a service by the Leibniz-Institute for Astrophysics Potsdam (AIP).
The MultiDark database was developed in cooperation with the Spanish MultiDark Consolider Project CSD2009-00064.

The authors gratefully acknowledge the Gauss Centre for Supercomputing e.V. (www.gauss-centre.eu) and the Partnership for Advanced Supercomputing in Europe (PRACE, www.prace-ri.eu) for funding the MultiDark simulation project by providing computing time on the GCS Supercomputer SuperMUC at Leibniz Supercomputing Centre (LRZ, www.lrz.de).
The Bolshoi simulations have been performed within the Bolshoi project of the University of California High-Performance AstroComputing Center (UC-HiPACC) and were run at the NASA Ames Research Center.

\bibliographystyle{mn2e}
\bibliography{Yuyu}

\begin{thebibliography}{}

\bibitem[\protect\citeauthoryear{{Abate} \& {Erdo{\v g}du}}{{Abate} \& {Erdo{\v
  g}du}}{2009}]{AbaErd2009}
{Abate} A.,  {Erdo{\v g}du} P.,  2009, \mnras, 400, 1541

\bibitem[\protect\citeauthoryear{{Abate} \& {Feldman}}{{Abate} \&
  {Feldman}}{2012}]{AbaFel2012}
{Abate} A.,  {Feldman} H.~A.,  2012, \mnras, 419, 3482

\bibitem[\protect\citeauthoryear{{Agarwal}, {Abdalla}, {Feldman}, {Lahav} \&
  {Thomas}}{{Agarwal} et~al.}{2012}]{AgaAbdFel2012}
{Agarwal} S.,  {Abdalla} F.~B.,  {Feldman} H.~A.,  {Lahav} O.,    {Thomas}
  S.~A.,  2012, \mnras, 424, 1409

\bibitem[\protect\citeauthoryear{{Agarwal}, {Abdalla}, {Feldman}, {Lahav} \&
  {Thomas}}{{Agarwal} et~al.}{2014}]{AgaAbdFel2014}
{Agarwal} S.,  {Abdalla} F.~B.,  {Feldman} H.~A.,  {Lahav} O.,    {Thomas}
  S.~A.,  2014, \mnras, 439, 2102

\bibitem[\protect\citeauthoryear{{Agarwal}, {Feldman} \& {Watkins}}{{Agarwal}
  et~al.}{2012}]{AgaFelWat2012}
{Agarwal} S.,  {Feldman} H.~A.,    {Watkins} R.,  2012, \mnras, 424, 2667

\bibitem[\protect\citeauthoryear{{Bennett}, {Larson}, {Weiland}, {Jarosik},
  {Hinshaw}, {Odegard}, {Gold}, {Halpern}, {Komatsu}, {Nolta}, {Page},
  {Spergel}, {Wollack}, {Dunkley}, {Kogut}, {Limon}, {Meyer}, {Tucker} \&
  {Wright}}{{Bennett} et~al.}{2013}]{wmap9}
{Bennett} C.~L.,  {Larson} D.,  {Weiland} J.~L.,  {Jarosik} N.,  {Hinshaw} G.,
  {Odegard} N.,  {Gold} B.,  {Halpern} M.,  {Komatsu} E.,  {Nolta} M.~R.,
  {Page} L.,  {Spergel} D.~N.,  {Wollack} E.,  {Dunkley} J.,  {Kogut} A.,
  {Limon} M.,  {Meyer} S.~S.,  {Tucker} G.~S.,    {Wright} E.~L.,  2013, \apjs,
  208, 20

\bibitem[\protect\citeauthoryear{{Bernardi}, {Alonso}, {da Costa}, {Willmer},
  {Wegner}, {Pellegrini}, {Rit{\'e}} \& {Maia}}{{Bernardi}
  et~al.}{2002}]{BerAlodaC2002}
{Bernardi} M.,  {Alonso} M.~V.,  {da Costa} L.~N.,  {Willmer} C.~N.~A.,
  {Wegner} G.,  {Pellegrini} P.~S.,  {Rit{\'e}} C.,    {Maia} M.~A.~G.,  2002,
  \aj, 123, 2990

\bibitem[\protect\citeauthoryear{{Bertschinger} \& {Dekel}}{{Bertschinger} \&
  {Dekel}}{1989}]{BerDek1989}
{Bertschinger} E.,  {Dekel} A.,  1989, \apjl, 336, L5

\bibitem[\protect\citeauthoryear{{Blobel}}{{Blobel}}{2003}]{Blobel2003}
{Blobel} V.,  2003, in {Lyons} L.,  {Mount} R.,   {Reitmeyer} R.,  eds,
  Statistical Problems in Particle Physics, Astrophysics, and Cosmology {Some
  Comments on X$^{2}$ Minimization Applications}.
p.~101

\bibitem[\protect\citeauthoryear{{Borgani}, {da Costa}, {Zehavi}, {Giovanelli},
  {Haynes}, {Freudling}, {Wegner} \& {Salzer}}{{Borgani}
  et~al.}{2000}]{BorCosZeh2000}
{Borgani} S.,  {da Costa} L.~N.,  {Zehavi} I.,  {Giovanelli} R.,  {Haynes}
  M.~P.,  {Freudling} W.,  {Wegner} G.,    {Salzer} J.~J.,  2000, \aj, 119, 102

\bibitem[\protect\citeauthoryear{{Colless}, {Saglia}, {Burstein}, {Davies},
  {McMahan} \& {Wegner}}{{Colless} et~al.}{2001}]{ColSagBur2001}
{Colless} M.,  {Saglia} R.~P.,  {Burstein} D.,  {Davies} R.~L.,  {McMahan}
  R.~K.,    {Wegner} G.,  2001, \mnras, 321, 277

\bibitem[\protect\citeauthoryear{{da Costa}, {Bernardi}, {Alonso}, {Wegner},
  {Willmer}, {Pellegrini}, {Rit{\'e}} \& {Maia}}{{da Costa}
  et~al.}{2000}]{daCBerAlo2000}
{da Costa} L.~N.,  {Bernardi} M.,  {Alonso} M.~V.,  {Wegner} G.,  {Willmer}
  C.~N.~A.,  {Pellegrini} P.~S.,  {Rit{\'e}} C.,    {Maia} M.~A.~G.,  2000,
  \aj, 120, 95

\bibitem[\protect\citeauthoryear{{D'Agostini}}{{D'Agostini}}{1995}]{DAgostini1995}
{D'Agostini} G.,  1995, ArXiv High Energy Physics - Phenomenology e-prints

\bibitem[\protect\citeauthoryear{{Dale}, {Giovanelli}, {Haynes}, {Campusano} \&
  {Hardy}}{{Dale} et~al.}{1999}]{DalGioHay1999}
{Dale} D.~A.,  {Giovanelli} R.,  {Haynes} M.~P.,  {Campusano} L.~E.,    {Hardy}
  E.,  1999, \aj, 118, 1489

\bibitem[\protect\citeauthoryear{{Davis}, {Nusser}, {Masters}, {Springob},
  {Huchra} \& {Lemson}}{{Davis} et~al.}{2011}]{DavNusMas2011}
{Davis} M.,  {Nusser} A.,  {Masters} K.~L.,  {Springob} C.,  {Huchra} J.~P.,
  {Lemson} G.,  2011, \mnras, 413, 2906

\bibitem[\protect\citeauthoryear{{Davis}, {Nusser} \& {Willick}}{{Davis}
  et~al.}{1996}]{DavNusWil1996}
{Davis} M.,  {Nusser} A.,    {Willick} J.~A.,  1996, \apj, 473, 22

\bibitem[\protect\citeauthoryear{{Davis} \& {Scrimgeour}}{{Davis} \&
  {Scrimgeour}}{2014}]{DavScr2014}
{Davis} T.~M.,  {Scrimgeour} M.~I.,  2014, \mnras, 442, 1117

\bibitem[\protect\citeauthoryear{{De Lucia} \& {Blaizot}}{{De Lucia} \&
  {Blaizot}}{2007}]{DeLBla2007}
{De Lucia} G.,  {Blaizot} J.,  2007, \mnras, 375, 2

\bibitem[\protect\citeauthoryear{{Djorgovski} \& {Davis}}{{Djorgovski} \&
  {Davis}}{1987}]{DjoDav1987}
{Djorgovski} S.,  {Davis} M.,  1987, \apj, 313, 59

\bibitem[\protect\citeauthoryear{{Dressler}, {Lynden-Bell}, {Burstein},
  {Davies}, {Faber}, {Terlevich} \& {Wegner}}{{Dressler}
  et~al.}{1987}]{DreLynBurDav1987}
{Dressler} A.,  {Lynden-Bell} D.,  {Burstein} D.,  {Davies} R.~L.,  {Faber}
  S.~M.,  {Terlevich} R.,    {Wegner} G.,  1987, \apj, 313, 42

\bibitem[\protect\citeauthoryear{{Eisenstein} \& {Hu}}{{Eisenstein} \&
  {Hu}}{1998}]{EisHu1998}
{Eisenstein} D.~J.,  {Hu} W.,  1998, \apj, 496, 605

\bibitem[\protect\citeauthoryear{{Faber} \& {Jackson}}{{Faber} \&
  {Jackson}}{1976}]{FaberJackson1976}
{Faber} S.~M.,  {Jackson} R.~E.,  1976, \apj, 204, 668

\bibitem[\protect\citeauthoryear{{Feldman}, {Juszkiewicz}, {Ferreira}, {Davis},
  {Gazta{\~n}aga}, {Fry}, {Jaffe}, {Chambers}, {da Costa}, {Bernardi},
  {Giovanelli}, {Haynes} \& {Wegner}}{{Feldman} et~al.}{2003}]{FelJusFer2003}
{Feldman} H.,  {Juszkiewicz} R.,  {Ferreira} P.,  {Davis} M.,  {Gazta{\~n}aga}
  E.,  {Fry} J.,  {Jaffe} A.,  {Chambers} S.,  {da Costa} L.,  {Bernardi} M.,
  {Giovanelli} R.,  {Haynes} M.,    {Wegner} G.,  2003, \apjl, 596, L131

\bibitem[\protect\citeauthoryear{{Feldman}, {Kaiser} \& {Peacock}}{{Feldman}
  et~al.}{1994}]{FKP}
{Feldman} H.~A.,  {Kaiser} N.,    {Peacock} J.~A.,  1994, \apj, 426, 23

\bibitem[\protect\citeauthoryear{{Feldman} \& {Watkins}}{{Feldman} \&
  {Watkins}}{2008}]{FelWat2008}
{Feldman} H.~A.,  {Watkins} R.,  2008, \mnras, 387, 825

\bibitem[\protect\citeauthoryear{{Feldman}, {Watkins} \& {Hudson}}{{Feldman}
  et~al.}{2010}]{FelWatHud2010}
{Feldman} H.~A.,  {Watkins} R.,    {Hudson} M.~J.,  2010, \mnras, 407, 2328

\bibitem[\protect\citeauthoryear{{Ferreira}, {Juszkiewicz}, {Feldman}, {Davis}
  \& {Jaffe}}{{Ferreira} et~al.}{1999}]{FerJusFel1999}
{Ferreira} P.~G.,  {Juszkiewicz} R.,  {Feldman} H.~A.,  {Davis} M.,    {Jaffe}
  A.~H.,  1999, \apjl, 515, L1

\bibitem[\protect\citeauthoryear{{Giovanelli}, {Haynes}, {Salzer}, {Wegner},
  {da Costa} \& {Freudling}}{{Giovanelli} et~al.}{1998}]{GioHaySal1998}
{Giovanelli} R.,  {Haynes} M.~P.,  {Salzer} J.~J.,  {Wegner} G.,  {da Costa}
  L.~N.,    {Freudling} W.,  1998, \aj, 116, 2632

\bibitem[\protect\citeauthoryear{{Gorski}}{{Gorski}}{1988}]{Gorski1988}
{Gorski} K.,  1988, \apjl, 332, L7

\bibitem[\protect\citeauthoryear{{Gorski}, {Davis}, {Strauss}, {White} \&
  {Yahil}}{{Gorski} et~al.}{1989}]{GorDavStr1989}
{Gorski} K.~M.,  {Davis} M.,  {Strauss} M.~A.,  {White} S.~D.~M.,    {Yahil}
  A.,  1989, \apj, 344, 1

\bibitem[\protect\citeauthoryear{{Guo}, {White}, {Angulo}, {Henriques},
  {Lemson}, {Boylan-Kolchin}, {Thomas} \& {Short}}{{Guo}
  et~al.}{2013}]{GuoWhiAng2013}
{Guo} Q.,  {White} S.,  {Angulo} R.~E.,  {Henriques} B.,  {Lemson} G.,
  {Boylan-Kolchin} M.,  {Thomas} P.,    {Short} C.,  2013, \mnras, 428, 1351

\bibitem[\protect\citeauthoryear{{Guzzo}, {Pierleoni}, {Meneux}, {Branchini},
  {Le F{\`e}vre}, {Marinoni}, {Garilli}, {Blaizot}, {De Lucia}, {Pollo},
  {McCracken} \& {Bottini}}{{Guzzo} et~al.}{2008}]{GuzPieMen2008}
{Guzzo} L.,  {Pierleoni} M.,  {Meneux} B.,  {Branchini} E.,  {Le F{\`e}vre} O.,
   {Marinoni} C.,  {Garilli} B.,  {Blaizot} J.,  {De Lucia} G.,  {Pollo} A.,
  {McCracken} H.~J.,    {Bottini} D.,  2008, \nat, 451, 541

\bibitem[\protect\citeauthoryear{{Hand}, {Addison}, {Aubourg}, {Battaglia},
  {Battistelli}, {Bizyaev}, {Bond}, {Brewington}, {Brinkmann}, {Brown}, {Das}
  \& {Dawson}}{{Hand} et~al.}{2012}]{HanAddAub2012}
{Hand} N.,  {Addison} G.~E.,  {Aubourg} E.,  {Battaglia} N.,  {Battistelli}
  E.~S.,  {Bizyaev} D.,  {Bond} J.~R.,  {Brewington} H.,  {Brinkmann} J.,
  {Brown} B.~R.,  {Das} S.,    {Dawson} K.~S.,  2012, Physical Review Letters,
  109, 041101

\bibitem[\protect\citeauthoryear{{Heitmann}, {Lawrence}, {Kwan}, {Habib} \&
  {Higdon}}{{Heitmann} et~al.}{2014}]{HeiLawKwaHab2014}
{Heitmann} K.,  {Lawrence} E.,  {Kwan} J.,  {Habib} S.,    {Higdon} D.,  2014,
  \apj, 780, 111

\bibitem[\protect\citeauthoryear{{Heitmann}, {White}, {Wagner}, {Habib} \&
  {Higdon}}{{Heitmann} et~al.}{2010}]{HeiWhiWagHab2010}
{Heitmann} K.,  {White} M.,  {Wagner} C.,  {Habib} S.,    {Higdon} D.,  2010,
  \apj, 715, 104

\bibitem[\protect\citeauthoryear{{Hellwing}}{{Hellwing}}{2014}]{Hellwing2014}
{Hellwing} W.~A.,  2014, ArXiv e-prints

\bibitem[\protect\citeauthoryear{{Hellwing}, {Barreira}, {Frenk}, {Li} \&
  {Cole}}{{Hellwing} et~al.}{2014}]{HelBarFre2014}
{Hellwing} W.~A.,  {Barreira} A.,  {Frenk} C.~S.,  {Li} B.,    {Cole} S.,
  2014, Physical Review Letters, 112, 221102

\bibitem[\protect\citeauthoryear{{Hellwing}, {Nusser}, {Feix} \&
  {Bilicki}}{{Hellwing} et~al.}{2017}]{HelNusFei2017}
{Hellwing} W.~A.,  {Nusser} A.,  {Feix} M.,    {Bilicki} M.,  2017, \mnras,
  467, 2787

\bibitem[\protect\citeauthoryear{{Hoffman}, {Nusser}, {Courtois} \&
  {Tully}}{{Hoffman} et~al.}{2016}]{HofNusCorTul2016}
{Hoffman} Y.,  {Nusser} A.,  {Courtois} H.~M.,    {Tully} R.~B.,  2016, ArXiv
  e-prints

\bibitem[\protect\citeauthoryear{{Howlett}, {Staveley-Smith} \&
  {Blake}}{{Howlett} et~al.}{2017}]{HowStaBla2017}
{Howlett} C.,  {Staveley-Smith} L.,    {Blake} C.,  2017, \mnras, 464, 2517

\bibitem[\protect\citeauthoryear{{Hudson}, {Smith}, {Lucey} \&
  {Branchini}}{{Hudson} et~al.}{2004}]{HudSmiLuc2004}
{Hudson} M.~J.,  {Smith} R.~J.,  {Lucey} J.~R.,    {Branchini} E.,  2004,
  \mnras, 352, 61

\bibitem[\protect\citeauthoryear{{Hudson}, {Smith}, {Lucey}, {Schlegel} \&
  {Davies}}{{Hudson} et~al.}{1999}]{HudSmiLuc1999}
{Hudson} M.~J.,  {Smith} R.~J.,  {Lucey} J.~R.,  {Schlegel} D.~J.,    {Davies}
  R.~L.,  1999, \apjl, 512, L79

\bibitem[\protect\citeauthoryear{{Jaffe} \& {Kaiser}}{{Jaffe} \&
  {Kaiser}}{1995}]{JafKai1995}
{Jaffe} A.~H.,  {Kaiser} N.,  1995, \apj, 455, 26

\bibitem[\protect\citeauthoryear{{Johnson}, {Blake}, {Koda}, {Ma}, {Colless},
  {Crocce}, {Davis}, {Jones}, {Magoulas}, {Lucey}, {Mould}, {Scrimgeour} \&
  {Springob}}{{Johnson} et~al.}{2014}]{JohBlaKod2014}
{Johnson} A.,  {Blake} C.,  {Koda} J.,  {Ma} Y.-Z.,  {Colless} M.,  {Crocce}
  M.,  {Davis} T.~M.,  {Jones} H.,  {Magoulas} C.,  {Lucey} J.~R.,  {Mould} J.,
   {Scrimgeour} M.~I.,    {Springob} C.~M.,  2014, \mnras, 444, 3926

\bibitem[\protect\citeauthoryear{{Juszkiewicz}, {Feldman}, {Fry} \&
  {Jaffe}}{{Juszkiewicz} et~al.}{2010}]{JusFelFryJaf2010}
{Juszkiewicz} R.,  {Feldman} H.~A.,  {Fry} J.~N.,    {Jaffe} A.~H.,  2010,
  \jcap, 2, 021

\bibitem[\protect\citeauthoryear{{Juszkiewicz}, {Ferreira}, {Feldman}, {Jaffe}
  \& {Davis}}{{Juszkiewicz} et~al.}{2000}]{JusFerFelJaf2000}
{Juszkiewicz} R.,  {Ferreira} P.~G.,  {Feldman} H.~A.,  {Jaffe} A.~H.,
  {Davis} M.,  2000, Science, 287, 109

\bibitem[\protect\citeauthoryear{{Kaiser}}{{Kaiser}}{1988}]{Kaiser1988}
{Kaiser} N.,  1988, \mnras, 231, 149

\bibitem[\protect\citeauthoryear{{Kaiser}}{{Kaiser}}{1989}]{Kaiser1989}
{Kaiser} N.,  1989, in {Mezzetti} M.,  {Giuricin} G.,  {Mardirossian} F.,
  {Ramella} M.,  eds, Large Scale Structure and Motions in the Universe
  Vol.~151 of Astrophysics and Space Science Library, {Local large scale
  structure VS cold dark matter}.
pp 197--212

\bibitem[\protect\citeauthoryear{{Kumar}, {Wang}, {Feldman} \&
  {Watkins}}{{Kumar} et~al.}{2015}]{KumWanFelWat2015}
{Kumar} A.,  {Wang} Y.,  {Feldman} H.~A.,    {Watkins} R.,  2015, ArXiv
  e-prints

\bibitem[\protect\citeauthoryear{{Larson}, {Dunkley}, {Hinshaw}, {Komatsu},
  {Nolta}, {Bennett}, {Gold}, {Halpern}, {Hill} \& {Jarosik}}{{Larson}
  et~al.}{2011}]{WMAP7}
{Larson} D.,  {Dunkley} J.,  {Hinshaw} G.,  {Komatsu} E.,  {Nolta} M.~R.,
  {Bennett} C.~L.,  {Gold} B.,  {Halpern} M.,  {Hill} R.~S.,    {Jarosik} 2011,
  \apjs, 192, 16

\bibitem[\protect\citeauthoryear{{Linder}}{{Linder}}{2005}]{Linder2005}
{Linder} E.~V.,  2005, \prd, 72, 043529

\bibitem[\protect\citeauthoryear{{Macaulay}, {Feldman}, {Ferreira}, {Hudson} \&
  {Watkins}}{{Macaulay} et~al.}{2011}]{MacFelFer2011}
{Macaulay} E.,  {Feldman} H.,  {Ferreira} P.~G.,  {Hudson} M.~J.,    {Watkins}
  R.,  2011, \mnras, 414, 621

\bibitem[\protect\citeauthoryear{{Macaulay}, {Feldman}, {Ferreira}, {Jaffe},
  {Agarwal}, {Hudson} \& {Watkins}}{{Macaulay} et~al.}{2012}]{MacFelFer2012}
{Macaulay} E.,  {Feldman} H.~A.,  {Ferreira} P.~G.,  {Jaffe} A.~H.,  {Agarwal}
  S.,  {Hudson} M.~J.,    {Watkins} R.,  2012, \mnras, 425, 1709

\bibitem[\protect\citeauthoryear{{Masters}, {Springob}, {Haynes} \&
  {Giovanelli}}{{Masters} et~al.}{2006}]{MasSprHay2006}
{Masters} K.~L.,  {Springob} C.~M.,  {Haynes} M.~P.,    {Giovanelli} R.,  2006,
  \apj, 653, 861

\bibitem[\protect\citeauthoryear{{Nusser}}{{Nusser}}{2014}]{Nusser2014}
{Nusser} A.,  2014, \apj, 795, 3

\bibitem[\protect\citeauthoryear{{Nusser}}{{Nusser}}{2016}]{Nusser2016}
{Nusser} A.,  2016, \mnras, 455, 178

\bibitem[\protect\citeauthoryear{{Nusser}, {Branchini} \& {Davis}}{{Nusser}
  et~al.}{2011}]{NusBraDav2011}
{Nusser} A.,  {Branchini} E.,    {Davis} M.,  2011, \apj, 735, 77

\bibitem[\protect\citeauthoryear{{Nusser} \& {Davis}}{{Nusser} \&
  {Davis}}{2011}]{NusDav2011}
{Nusser} A.,  {Davis} M.,  2011, \apj, 736, 93

\bibitem[\protect\citeauthoryear{{Okumura}, {Seljak}, {Vlah} \&
  {Desjacques}}{{Okumura} et~al.}{2014}]{OkuSelVla2014}
{Okumura} T.,  {Seljak} U.,  {Vlah} Z.,    {Desjacques} V.,  2014, \jcap, 5,
  003

\bibitem[\protect\citeauthoryear{{Planck Collaboration}}{{Planck
  Collaboration}}{2014}]{Planckparameters2014}
{Planck Collaboration} 2014, \aap, 571, A16

\bibitem[\protect\citeauthoryear{{Planck Collaboration}}{{Planck
  Collaboration}}{2016}]{PlanckXXXVII2015}
{Planck Collaboration} 2016, \aap, 586, A140

\bibitem[\protect\citeauthoryear{{Riess}, {Macri}, {Hoffmann}, {Scolnic},
  {Casertano}, {Filippenko}, {Tucker}, {Reid}, {Jones}, {Silverman},
  {Chornock}, {Challis}, {Yuan}, {Brown} \& {Foley}}{{Riess}
  et~al.}{2016}]{RieMacHof2016}
{Riess} A.~G.,  {Macri} L.~M.,  {Hoffmann} S.~L.,  {Scolnic} D.,  {Casertano}
  S.,  {Filippenko} A.~V.,  {Tucker} B.~E.,  {Reid} M.~J.,  {Jones} D.~O.,
  {Silverman} J.~M.,  {Chornock} R.,  {Challis} P.,  {Yuan} W.,  {Brown} P.~J.,
     {Foley} R.~J.,  2016, \apj, 826, 56

\bibitem[\protect\citeauthoryear{{Scrimgeour}, {Davis}, {Blake},
  {Staveley-Smith}, {Magoulas}, {Springob}, {Beutler}, {Colless}, {Johnson},
  {Jones}, {Koda}, {Lucey}, {Ma}, {Mould} \& {Poole}}{{Scrimgeour}
  et~al.}{2016a}]{ScrDavBlaSta2015}
{Scrimgeour} M.~I.,  {Davis} T.~M.,  {Blake} C.,  {Staveley-Smith} L.,
  {Magoulas} C.,  {Springob} C.~M.,  {Beutler} F.,  {Colless} M.,  {Johnson}
  A.,  {Jones} D.~H.,  {Koda} J.,  {Lucey} J.~R.,  {Ma} Y.-Z.,  {Mould} J.,
  {Poole} G.~B.,  2016a, \mnras, 455, 386

\bibitem[\protect\citeauthoryear{{Scrimgeour}, {Davis}, {Blake},
  {Staveley-Smith}, {Magoulas}, {Springob}, {Beutler}, {Colless}, {Johnson},
  {Jones}, {Koda}, {Lucey}, {Ma}, {Mould} \& {Poole}}{{Scrimgeour}
  et~al.}{2016b}]{ScrDavBla2015}
{Scrimgeour} M.~I.,  {Davis} T.~M.,  {Blake} C.,  {Staveley-Smith} L.,
  {Magoulas} C.,  {Springob} C.~M.,  {Beutler} F.,  {Colless} M.,  {Johnson}
  A.,  {Jones} D.~H.,  {Koda} J.,  {Lucey} J.~R.,  {Ma} Y.-Z.,  {Mould} J.,
  {Poole} G.~B.,  2016b, \mnras, 455, 386

\bibitem[\protect\citeauthoryear{{Seiler} \& {Parkinson}}{{Seiler} \&
  {Parkinson}}{2016}]{SeiPar2016}
{Seiler} J.,  {Parkinson} D.,  2016, \mnras, 462, 75

\bibitem[\protect\citeauthoryear{{Springel}, {White}, {Jenkins}, {Frenk},
  {Yoshida}, {Gao}, {Navarro}, {Thacker}, {Croton}, {Helly}, {Peacock}, {Cole},
  {Thomas}, {Couchman}, {Evrard}, {Colberg} \& {Pearce}}{{Springel}
  et~al.}{2005}]{Millennium1}
{Springel} V.,  {White} S.~D.~M.,  {Jenkins} A.,  {Frenk} C.~S.,  {Yoshida} N.,
   {Gao} L.,  {Navarro} J.,  {Thacker} R.,  {Croton} D.,  {Helly} J.,
  {Peacock} J.~A.,  {Cole} S.,  {Thomas} P.,  {Couchman} H.,  {Evrard} A.,
  {Colberg} J.,    {Pearce} F.,  2005, \nat, 435, 629

\bibitem[\protect\citeauthoryear{{Springob}, {Magoulas}, {Colless}, {Mould},
  {Erdo{\u g}du}, {Jones}, {Lucey}, {Campbell} \& {Fluke}}{{Springob}
  et~al.}{2014a}]{SprMagColMou2014}
{Springob} C.~M.,  {Magoulas} C.,  {Colless} M.,  {Mould} J.,  {Erdo{\u g}du}
  P.,  {Jones} D.~H.,  {Lucey} J.~R.,  {Campbell} L.,    {Fluke} C.~J.,  2014a,
  \mnras, 445, 2677

\bibitem[\protect\citeauthoryear{{Springob}, {Magoulas}, {Colless}, {Mould},
  {Erdo{\u g}du}, {Jones}, {Lucey}, {Campbell} \& {Fluke}}{{Springob}
  et~al.}{2014b}]{SprMagCol2014}
{Springob} C.~M.,  {Magoulas} C.,  {Colless} M.,  {Mould} J.,  {Erdo{\u g}du}
  P.,  {Jones} D.~H.,  {Lucey} J.~R.,  {Campbell} L.,    {Fluke} C.~J.,  2014b,
  \mnras, 445, 2677

\bibitem[\protect\citeauthoryear{{Springob}, {Masters}, {Haynes}, {Giovanelli}
  \& {Marinoni}}{{Springob} et~al.}{2007}]{SprMasHay2007}
{Springob} C.~M.,  {Masters} K.~L.,  {Haynes} M.~P.,  {Giovanelli} R.,
  {Marinoni} C.,  2007, \apjs, 172, 599

\bibitem[\protect\citeauthoryear{{Springob}, {Masters}, {Haynes}, {Giovanelli}
  \& {Marinoni}}{{Springob} et~al.}{2009}]{SprMasHay2009}
{Springob} C.~M.,  {Masters} K.~L.,  {Haynes} M.~P.,  {Giovanelli} R.,
  {Marinoni} C.,  2009, \apjs, 182, 474

\bibitem[\protect\citeauthoryear{{Tonry}, {Dressler}, {Blakeslee}, {Ajhar},
  {Fletcher}, {Luppino}, {Metzger} \& {Moore}}{{Tonry}
  et~al.}{2001}]{TonDreBla2001}
{Tonry} J.~L.,  {Dressler} A.,  {Blakeslee} J.~P.,  {Ajhar} E.~A.,  {Fletcher}
  A.~B.,  {Luppino} G.~A.,  {Metzger} M.~R.,    {Moore} C.~B.,  2001, \apj,
  546, 681

\bibitem[\protect\citeauthoryear{{Tonry}, {Schmidt}, {Barris}, {Candia},
  {Challis}, {Clocchiatti}, {Coil}, {Filippenko}, {Garnavich}, {Hogan},
  {Holland}, {Jha}, {Kirshner}, {Krisciunas}, {Leibundgut}, {Li} \&
  {Matheson}}{{Tonry} et~al.}{2003}]{TonSchBar}
{Tonry} J.~L.,  {Schmidt} B.~P.,  {Barris} B.,  {Candia} P.,  {Challis} P.,
  {Clocchiatti} A.,  {Coil} A.~L.,  {Filippenko} A.~V.,  {Garnavich} P.,
  {Hogan} C.,  {Holland} S.~T.,  {Jha} S.,  {Kirshner} R.~P.,  {Krisciunas} K.,
   {Leibundgut} B.,  {Li} W.,    {Matheson} T.,  2003, \apj, 594, 1

\bibitem[\protect\citeauthoryear{{Tully}, {Courtois}, {Dolphin}, {Fisher},
  {H{\'e}raudeau}, {Jacobs}, {Karachentsev}, {Makarov}, {Makarova},
  {Mitronova}, {Rizzi}, {Shaya}, {Sorce} \& {Wu}}{{Tully}
  et~al.}{2013}]{TulCouDol2013}
{Tully} R.~B.,  {Courtois} H.~M.,  {Dolphin} A.~E.,  {Fisher} J.~R.,
  {H{\'e}raudeau} P.,  {Jacobs} B.~A.,  {Karachentsev} I.~D.,  {Makarov} D.,
  {Makarova} L.,  {Mitronova} S.,  {Rizzi} L.,  {Shaya} E.~J.,  {Sorce} J.~G.,
    {Wu} P.-F.,  2013, \aj, 146, 86

\bibitem[\protect\citeauthoryear{{Tully}, {Courtois} \& {Sorce}}{{Tully}
  et~al.}{2016}]{TulCouSor2016}
{Tully} R.~B.,  {Courtois} H.~M.,    {Sorce} J.~G.,  2016, \aj, 152, 50

\bibitem[\protect\citeauthoryear{{Tully} \& {Fisher}}{{Tully} \&
  {Fisher}}{1977}]{TullyFisher1977}
{Tully} R.~B.,  {Fisher} J.~R.,  1977, \aap, 54, 661

\bibitem[\protect\citeauthoryear{{Turnbull}, {Hudson}, {Feldman}, {Hicken},
  {Kirshner} \& {Watkins}}{{Turnbull} et~al.}{2012}]{Turnbull2012}
{Turnbull} S.~J.,  {Hudson} M.~J.,  {Feldman} H.~A.,  {Hicken} M.,  {Kirshner}
  R.~P.,    {Watkins} R.,  2012, \mnras, 420, 447

\bibitem[\protect\citeauthoryear{{Watkins} \& {Feldman}}{{Watkins} \&
  {Feldman}}{2007}]{WatFel2007}
{Watkins} R.,  {Feldman} H.~A.,  2007, \mnras, 379, 343

\bibitem[\protect\citeauthoryear{{Watkins} \& {Feldman}}{{Watkins} \&
  {Feldman}}{2015a}]{WatFel2015a}
{Watkins} R.,  {Feldman} H.~A.,  2015a, \mnras, 450, 1868

\bibitem[\protect\citeauthoryear{{Watkins} \& {Feldman}}{{Watkins} \&
  {Feldman}}{2015b}]{WatFel2015}
{Watkins} R.,  {Feldman} H.~A.,  2015b, \mnras, 447, 132

\bibitem[\protect\citeauthoryear{{Watkins}, {Feldman} \& {Hudson}}{{Watkins}
  et~al.}{2009}]{WatFelHud2009}
{Watkins} R.,  {Feldman} H.~A.,    {Hudson} M.~J.,  2009, \mnras, 392, 743

\bibitem[\protect\citeauthoryear{{Wegner}, {Bernardi}, {Willmer}, {da Costa},
  {Alonso}, {Pellegrini}, {Maia}, {Chaves} \& {Rit{\'e}}}{{Wegner}
  et~al.}{2003}]{WegBerWil2003}
{Wegner} G.,  {Bernardi} M.,  {Willmer} C.~N.~A.,  {da Costa} L.~N.,  {Alonso}
  M.~V.,  {Pellegrini} P.~S.,  {Maia} M.~A.~G.,  {Chaves} O.~L.,    {Rit{\'e}}
  C.,  2003, \aj, 126, 2268

\bibitem[\protect\citeauthoryear{{Willick}}{{Willick}}{1999}]{Willick1999}
{Willick} J.~A.,  1999, \apj, 516, 47

\bibitem[\protect\citeauthoryear{{Willick}, {Strauss}, {Dekel} \&
  {Kolatt}}{{Willick} et~al.}{1997}]{WilStrDek1997}
{Willick} J.~A.,  {Strauss} M.~A.,  {Dekel} A.,    {Kolatt} T.,  1997, \apj,
  486, 629

\bibitem[\protect\citeauthoryear{Wishart}{Wishart}{1928}]{Wishart}
Wishart J.,  1928, Biometrika, 20A, 32

\bibitem[\protect\citeauthoryear{{Zaroubi}, {Zehavi}, {Dekel}, {Hoffman} \&
  {Kolatt}}{{Zaroubi} et~al.}{1997}]{ZarZehDekHof1997}
{Zaroubi} S.,  {Zehavi} I.,  {Dekel} A.,  {Hoffman} Y.,    {Kolatt} T.,  1997,
  \apj, 486, 21

\bibitem[\protect\citeauthoryear{{Zhang}, {Feldman}, {Juszkiewicz} \&
  {Stebbins}}{{Zhang} et~al.}{2008}]{ZhaFelJus2008}
{Zhang} P.,  {Feldman} H.~A.,  {Juszkiewicz} R.,    {Stebbins} A.,  2008,
  \mnras, 388, 884

\end{thebibliography}

\end{document}